\documentclass[11pt,twoside,a4paper,pdf]{article}  
\usepackage{graphicx,color}
\usepackage{times}
\usepackage{xspace}
\usepackage{booktabs}
\usepackage[english]{babel}
\usepackage{subfig}
\usepackage{amssymb}
\usepackage{cite}
\usepackage{xspace}
\usepackage{a4wide}
\usepackage[colorlinks=true, pdftex, citecolor=red,
  pdftitle={mcplots.cern.ch}, pdfauthor={P. Skands et al.},
  pdfsubject={Monte Carlo generators}, pdfkeywords={Monte Carlo, event
    generators, QCD, HEP, Rivet, HEPDATA, validation, tuning},
    bookmarksnumbered=true,     
    bookmarksopen=true,         
    bookmarksopenlevel=1]{hyperref}

\newcommand{\mrm}[1]{\ensuremath{\mathrm{#1}}\xspace}
\newcommand{\ttt}[1]{\texttt{#1}}

\newcommand{\alp}{\textsc{alpgen}\xspace}
\newcommand{\mg}{\textsc{madgraph}\xspace}
\newcommand{\sherpa}{\textsc{sherpa}\xspace}
\newcommand{\py}{\textsc{pythia}\xspace}
\newcommand{\hw}{\textsc{herwig}\xspace}
\newcommand{\hwpp}{\textsc{herwig++}\xspace}
\newcommand{\sh}{\textsc{sherpa}\xspace}
\newcommand{\vc}{\textsc{vincia}\xspace}
\newcommand{\rivet}{\textsc{rivet}\xspace}
\newcommand{\cernvm}{\textsc{cernvm}\xspace}
\newcommand{\epos}{\textsc{epos}\xspace}
\newcommand{\hepmc}{\textsc{hepmc}\xspace}
\newcommand{\lhapdf}{\textsc{lhapdf}\xspace}
\newcommand{\fastjet}{\textsc{fastjet}\xspace}
\newcommand{\sixtrack}{\textsc{sixtrack}\xspace}
\newcommand{\agile}{\textsc{agile}\xspace}
\newcommand{\mcplots}{\textsc{mcplots}\xspace}
\newcommand{\pho}{\textsc{phojet}\xspace}
\newcommand{\tft}{\textsc{test4theory}\xspace}
\newcommand{\copilot}{\textsc{copilot}\xspace}
\newcommand{\lhcathome}{\textsc{lhc@home~2.0}\xspace}
\newcommand{\lhcathomeclassic}{\textsc{lhc@home}\xspace}
\newcommand{\hepdata}{\textsc{hepdata}\xspace}
\newcommand{\boinc}{\textsc{Boinc}\xspace}

\newcommand{\html}{{\footnotesize HTML}\xspace}
\newcommand{\latex}{{\footnotesize LaTeX}\xspace}

\newcommand{\code}[1]{{\small\ttt{#1}}\xspace}

\newcommand{\eqRef}[1]{equation~(\ref{#1})\xspace}

\newcommand{\secRef}[1]{section~\ref{#1}\xspace}
\newcommand{\SecRef}[1]{Section~\ref{#1}\xspace}
\newcommand{\secsRef}[1]{sections~\ref{#1}\xspace}

\newcommand{\figRef}[1]{figure~\ref{#1}\xspace}
\newcommand{\FigRef}[1]{Figure~\ref{#1}\xspace}

\newcommand{\inst}[1]{$^{#1}$}
\renewcommand{\and}{, }

\newenvironment{myitemize}{\begin{list}{$\diamond$}%
{\setlength{\topsep}{2mm}\setlength{\partopsep}{1mm}\setlength\leftmargin{6mm}%
\setlength{\itemsep}{0.3mm}\setlength{\parsep}{1mm}}}%
{\end{list}}

\newcommand{\vbspc}{2ex}

\begin{document}

\vspace*{-1.8cm}\begin{minipage}{\textwidth}
\flushright
CERN-PH-TH/2013-105\\
DESY 13-104\\
LU-TP 13-23\\
\end{minipage}
\vskip1.25cm
\begin{center}
{\Large\bf MCPLOTS: a particle physics resource 
based on volunteer computing}
\end{center}
\vskip5mm
{\begin{center}
{\large 
A.~ Karneyeu\inst{1}\and 
L.~Mijovic\inst{2,3}\and
S.~Prestel\inst{2,4}\and 
P.~Z.~Skands\inst{5}
}\end{center}\vskip3mm
\begin{center}
\parbox{0.9\textwidth}{
\inst{1}: Institute for Nuclear Research, Moscow, Russia \\
\inst{2}: Deutsches Elektronen-Synchrotron, DESY, 22603 Hamburg, Germany\\
\inst{3}: Irfu/SPP, CEA-Saclay, bat 141, F-91191 Gif-sur-Yvette Cedex, France\\
\inst{4}: Dept.\ of Astronomy and Theoretical Physics, Lund University, Sweden\\
\inst{5}: European Organization for Nuclear Research, CERN, CH-1211,
Geneva 23, Switzerland
}
\end{center}
\vskip5mm
\begin{center}
\parbox{0.85\textwidth}{
\begin{center}
\textbf{Abstract}
\end{center}\small
The {\bf mcplots.cern.ch} web site (\mcplots) provides 
a simple online repository of plots made with 
high-energy-physics event generators, comparing them to a wide
variety of experimental data. The repository is based on the \hepdata
online database of experimental results and on the \rivet Monte Carlo
analysis tool. The repository is continually updated and 
relies on computing power donated by volunteers, via the 
\lhcathome platform. 
}
\end{center}
\vspace*{1cm}
\section{Introduction \label{sec:intro}}

Computer simulations of high-energy interactions are used to
provide an explicit theoretical reference for a wide range of
particle-physics measurements. In particular, Monte Carlo (MC) 
event generators \cite{Buckley:2011ms,Skands:2012ts,Seymour:2013ega} enable a  
comparison between theory and experimental data down to the level of
individual particles. An exact calculation taking all relevant
dynamics into account would require a solution to 
infinite-order perturbation theory
coupled to non-perturbative QCD --- a long-standing and unsolved
problem. In the absence of such a solution,
MC generators apply a divide-and-conquer strategy, 
factorizing the problem into many simpler pieces, and 
treating each one at a level of approximation dictated by our
understanding of the corresponding parts of the underlying fundamental
theory.  

A central question, when a disagreement is found between simulation
and data, is thus whether the discrepancy is within the intrinsic 
uncertainty allowed by the inaccuracy of the
calculation, or not. This accuracy depends both on the sophistication of the
simulation itself, driven by the development and implementation of new
theoretical ideas, but it also depends crucially on the available
constraints on the free parameters of the model. Using existing data
to constrain these is referred to as
``tuning''. Useful discussions of tuning can be found, e.g., 
in~\cite{Buckley:2011ms,Buckley:2009bj,Skands:2010ak,Corke:2010yf,Gieseke:2008ep,Schulz:2011qy,ATLAS:2012uec,AlcarazMaestre:2012vp,Skands:2012ts}.   

Typically, experimental studies include comparisons of specific
models and tunes to the data in their publications. 
Such comparisons are useful both as immediate  tests of commonly used
models, and to illustrate the current amount of theoretical
uncertainty surrounding a particular distribution.  They also 
provide a set of well-defined theoretical reference 
curves that can be useful as benchmarks for 
future studies. However, 
many physics distributions, in particular those that are 
infrared (IR) sensitive\footnote{IR sensitive observables change value when
  adding an infinitely soft particle or when splitting an existing
  particle into two collinear ones. Such variables have larger
  sensitivity to non-perturbative physics than IR safe ones, see,
  e.g., \cite{Salam:2010zt,Skands:2012ts}. Note also that we here use
  the word ``IR'' as a catchall for both the soft and collinear
  limits. }
often represent a complicated cocktail of physics effects. 
The conclusions that can be drawn from comparisons on individual
distributions are therefore  limited. They also 
gradually become outdated, as new models and tunes supersede the old ones.
In the long term, the real aim
is not to study one distribution in detail, for which a
simple fit would in principle suffice, but to study the
degree of simultaneous agreement or disagreement over many, mutually
complementary, distributions. This is also a prerequisite to extend the
concept of tuning to more rigorous consistency tests of the underlying
physics model, for instance as proposed in \cite{Schulz:2011qy}. 

The effort involved in making
simultaneous comparisons to large
numbers of experimental distributions, and to keep those up to date,
previously meant that this kind of exercise was restricted 
to a small set of people, mostly Monte Carlo authors and dedicated
experimental tuning groups. The aim with the \textbf{mcplots.cern.ch}
(\mcplots)  
web site is to provide a simple
browsable repository of such comparisons 
so that anyone can quickly get an idea
of how well a particular model describes various data
sets\footnote{Note: this idea was 
  first raised in the now defunct \textsc{jetweb} project
  \cite{Butterworth:2002ts}. The first set of draft web pages for
  \mcplots\ were created following the 2005 Les Houches
  workshop~\cite{Skands:2007zz,Buttar:2008jx}. The CERN Generator 
  Services (GENSER) 
  project also maintains a set of
  \href{http://sftweb.cern.ch/generators}{web pages} with basic 
  {generator validation
    plots/calculations}.}.   
Simultaneously, we also
aim to make all generated data, 
parameter cards, source codes, experimental references, etc,
freely and directly 
available in as useful forms as possible, for anyone who wishes
to reproduce, re-plot, or otherwise make use of the results
and tools that we have developed.

The \mcplots web site is now at a mature and stable stage.
It has been online since Dec 2010 and is nearing a
trillion generated events in total (900 billion as of June 2013). 
This paper is intended to give an overview of what is available on the
site, and how to use it. In particular, \secRef{sec:userGuide}
contains a brief ``user guide'' for the site, explaining its features
and contents in simple terms, how to navigate through the site, and
how to extract plots and information about how they were made from
it. As a reference for further additions and updates, and for the
benefit of future developers, \secsRef{sec:architecture} --
\ref{sec:t4t} then describe the more concrete details of the
technical structure and implementation of the site, which the ordinary
user would not need to be familiar with. 

\SecRef{sec:architecture} describes the architecture of the site, and the thinking behind it. It currently relies on the following 
basic prerequisites, 
\begin{itemize} 
\item The \hepdata database \cite{Buckley:2010jn} 
  of experimental results.
\item The \rivet Monte Carlo analysis tool \cite{Buckley:2010ar} which contains
  a large code library for 
  comparing the output of MC generators to distributions from
  \hepdata. \rivet in turn relies on the \hepmc universal 
  event-record format \cite{Dobbs:2001ck}, on the \fastjet 
package for jet clustering 
\cite{Cacciari:2005hq,Cacciari:2011ma}, 
and on the \lhapdf library for parton densities
\cite{Whalley:2005nh,Bourilkov:2006cj}. 
\item Monte Carlo event generators. Currently implemented generators 
include \alp~\cite{Mangano:2002ea}, \epos~\cite{Werner:2010aa}
  \hwpp~\cite{Bahr:2008pv}, \pho~\cite{Bopp:1998rc},
  \py~6~\cite{Sjostrand:2006za}, 
  \py~8~\cite{Sjostrand:2007gs}, \sh~\cite{Gleisberg:2008ta}, 
and \vc~\cite{Giele:2011cb}. Some of these in turn use the 
Les Houches Event File (LHEF) format \cite{Boos:2001cv,Alwall:2006yp} to pass
parton-level information back and forth. 
\item The \lhcathome framework for volunteer cloud
  computing~\cite{lhcathome,Segal:2010zz,AguadoSanchez:2011zz,LombranaGonzalez:2012gd}. 
\lhcathome in turn relies on \cernvm~\cite{Buncic:2010zz,Harutyunyan:2012un} (a 
Virtual-Machine computing environment based on Scientific Linux), on the
\copilot job submission
system~\cite{Harutyunyan:2011zz,Harutyunyan:2012un}, and on the \boinc
middleware for volunteer computing~\cite{boinc,Hoimyr:2012ur}. 
\end{itemize}

The basic procedure to include a new measurement on
\mcplots is, first, to provide the relevant experimental 
data points to \hepdata,
second to provide a \rivet routine for the corresponding analysis, and
lastly to provide a very small additional amount of information to
\mcplots, essentially specifying the placement of the observable in the \mcplots
menus and summarizing the cuts applied in a \latex string, as
e.g.\ exemplified by the already existing analyses on the site. This
is described in more detail in \secRef{sec:newAnalysis}. 

To  update \mcplots with a new version of an existing generator,
 the first step is to check whether it is already 
available in the standard CERN 
Generator Services (GENSER) repository~\cite{genser}, 
and if not announce the new
version to the GENSER team. The \mcplots steering scripts
should then be updated to run jobs for the new version, as 
 described in \secRef{sec:newVersion}.

To add a new generator to \mcplots, the first step is to check that 
it can run within \cernvm. \cernvm
provides a standardized Scientific-Linux environment that
should be appropriate for most high-energy physics (HEP)
applications, including several commonly used auxiliary packages such
as the GNU Scientific Library (GSL), the C++ BOOST libraries, and many
others. A standalone version of \cernvm  
can be downloaded from
\cite{Buncic:2010zz} for testing purposes. To the extent that dependencies
require additional packages to be installed, these should be
communicated to the \mcplots and \cernvm development teams. The code
should then be provided to the GENSER team for inclusion in the
standard CERN generator repository. The complete procedure is
described in more detail in \secRef{sec:newGenerator}.

The main benefit of using \cernvm is that this 
has enabled \mcplots to draw on significant computing resources 
made available by volunteers, via the \boinc\ and \lhcathome\
projects. 
Through the
intermediary of \cernvm, 
generator authors can concentrate on developing code that is
compatible with the Scientific Linux operating system, a fairly
standard environment in our field. This code can 
then be run on essentially any user platform
by encapsulating it within \cernvm\ and distributing it via the 
\boinc\ middleware. (The latter also enabled us to access a large
existing \boinc\ volunteer-computing community.)
The resulting ``\tft'' project
\cite{test4theory,LombranaGonzalez:2012gd} was the first application 
developed for the \lhcathome framework,
see~\cite{lhcathome,LombranaGonzalez:2012gd}, and it 
represents the world's first virtualization-based volunteer cloud. A
brief summary of it is given in \secRef{sec:t4t}.

\section{User Guide \label{sec:userGuide}}

In this section, we describe the graphical interface on the \mcplots
web site and how to navigate through it. Care has been taken to design
it so as to make all content accessible through a few clicks, in a
hopefully intuitive manner.  

\subsection{The Main Menu}
\begin{figure}[h!t]
\centering
\includegraphics*[width=0.3\textwidth]{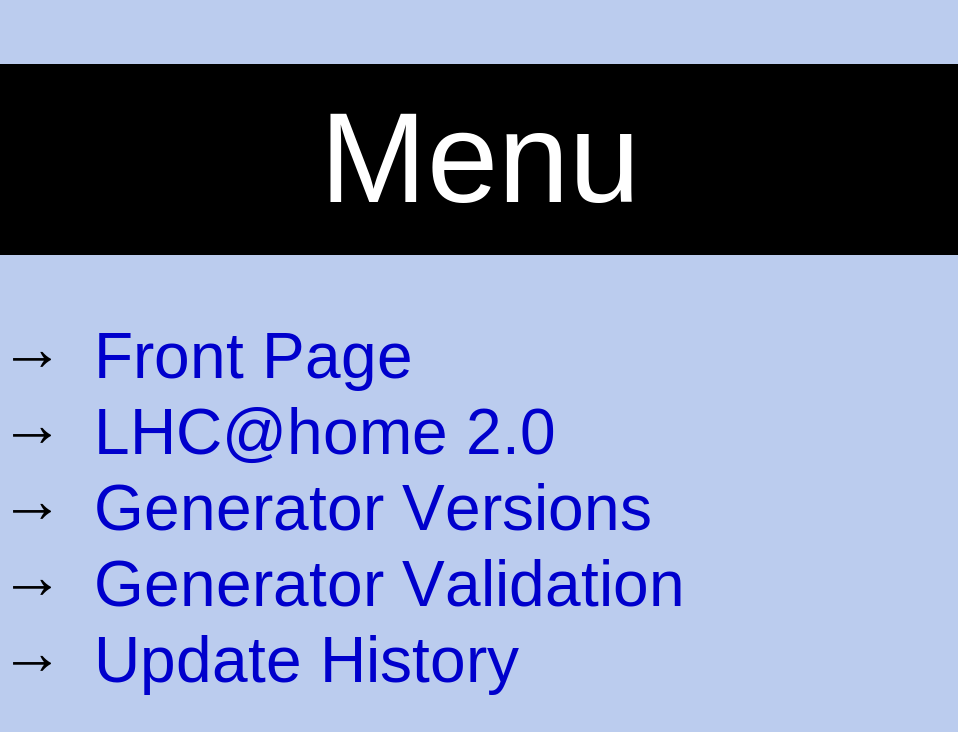} 
\caption{The main \mcplots menu.\label{fig:mainMenu}}
\end{figure}

The main menu, shown in \figRef{fig:mainMenu}, 
is always located at the top left-hand corner of the
page. The {\sl Front Page} link is just a ``home'' button that takes you
back to the starting page for \mcplots, and the {\sl LHC@home 2.0} one
takes you to the external \lhcathome web pages, where you can connect your
computer to the volunteer cloud that generates events for \mcplots. 

The {\sl Generator Versions} link opens a configuration page that allows you to
select which generator versions you want to see results for on the
site. The default is simply to use the most recently implemented ones, 
but if your analysis, for instance, uses an older version of
a particular generator, you can select among any of the previously
implemented versions on the site by choosing that specific version on 
the {\sl Generator Versions} page. All displayed content on the site
will thereafter reflect your choice, as you can verify by checking the
explicit version numbers written at the bottom of each plot. You can 
return to the {\sl Generator Versions} page at any time to modify your
choice. After making your choice, 
click on the {\sl Front Page} button to exit the {\sl Generator
  Versions} view. 

The {\sl Generator Validation} link changes the page layout
and content from the {\sl Front Page} one, to one in which different
generator versions can be compared both globally, via $\chi^2$
values, and individually on each distribution. This view will be
discussed in more detail in \secRef{sec:validationView}. 

The {\sl Update History} link simply takes you to a page on which you
can see what the most recent changes and additions to the site were,
and its previous history. 

As an experimental social feature, we have added a ``like'' button to the
bottom of the front page, which you can use to express if you are
happy with the \mcplots site. 

\subsection{The Analysis Filter} 
Immediately below the main menu, we have
collected a few options to control and organize which analyses you
want to see displayed on the site, under the subheading {\sl Analysis
  Filter}, illustrated in \figRef{fig:analysisFilter}. 

\begin{figure}[h!t]
\centering
\subfloat[Analysis Filter: Normal View]{
\parbox{0.4\textwidth}{\centering\includegraphics*[width=0.3\textwidth]{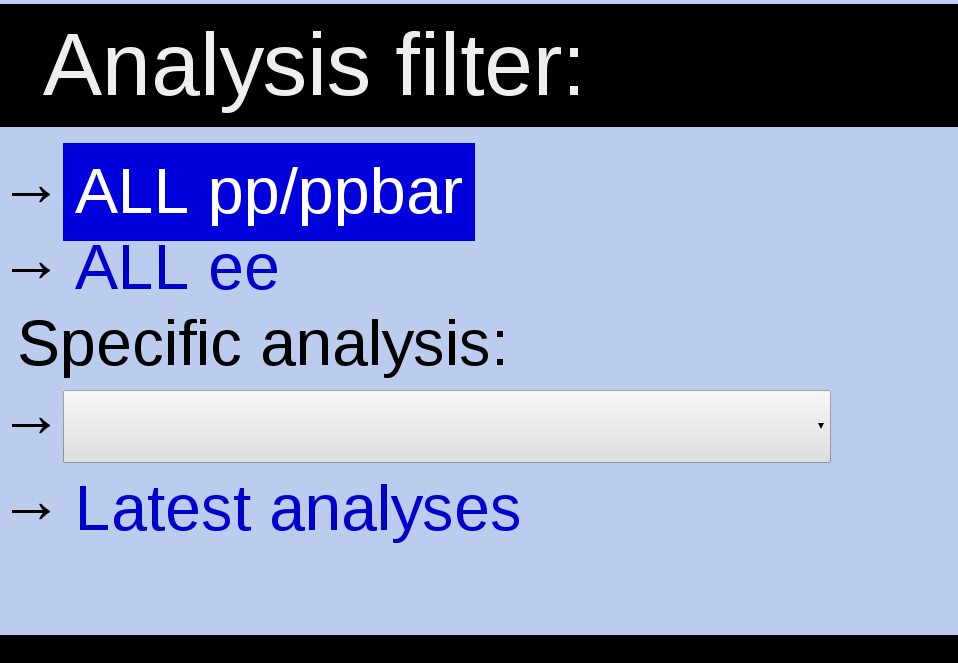}} }
\subfloat[Analysis Filter: Dropdown
  Menu \label{fig:specificAnalysis}]{\parbox{0.4\textwidth}{\centering\includegraphics*[width=0.3\textwidth]{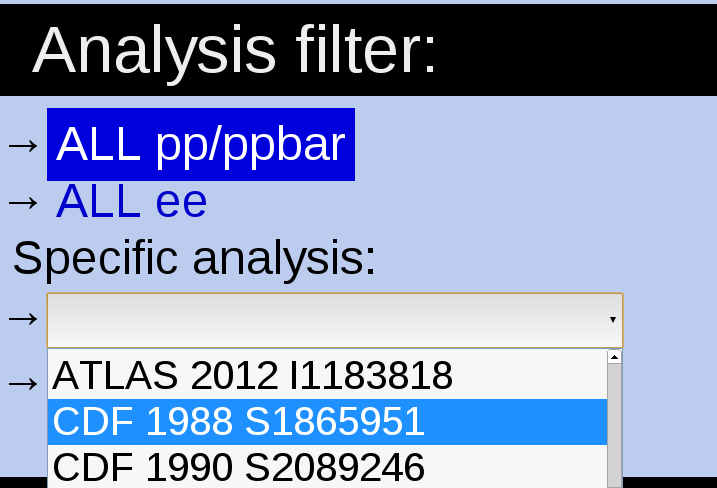}}} 
\caption{The analysis filter submenu; normal view (left) and after
  clicking on the {\sl
    Specific Analysis} dropdown menu (right).\label{fig:analysisFilter}}
\end{figure}

At the time of writing, the main choice you have here 
is between {\sl ALL pp/ppbar} (for hadron collisions) and {\sl
  ALL ee} (for fragmentation in electron-positron collisions). The
default is {\sl ALL pp/ppbar}, so if you are interested in seeing all
hadron-collider analyses, you will not have to make any changes here. 
Using
the {\sl Specific Analysis} dropdown menu, you can also select to see
only the plots from one particular \rivet analysis. The latter currently
requires that you know the \rivet ID of the analysis you are
interested in. The 
ID is typically formed from the experiment name, the year, and the
inSPIRE ID (or SPIRES ID, for older analyses) of the paper containing 
the original analysis, as illustrated in
\figRef{fig:specificAnalysis} (the numbers beginning with ``I'' are inSPIRE
codes, while ones beginning with ``S'' are SPIRES ones). 
You can also find this 
information in the \rivet user manual~\cite{Buckley:2010ar} and/or on 
the \href{http://rivet.hepforge.org/}{\rivet web pages}.

Finally, if you click on {\sl Latest Analyses}, only those analyses that were
added in the last update of the site will be shown. This can be
useful to get a quick overview of what is new on the site, for instance to
check for new distributions that are relevant
to you and that you may not have been able to see on your 
last visit to the site.
More options may of course be added in the future, in particular as the
number of observables added to the pp/ppbar set grows.

\subsection{Selecting Observables \label{subsec:SelectingObs}}

Below the {\sl Analysis Filter}, the list of processes and observables
for the selected set of analyses is shown. This is illustrated in
\figRef{fig:obsList}. 
\begin{figure}[h!t]
\centering
\subfloat[The Process and Observables List \label{fig:obsListDef}]{
\parbox{0.4\textwidth}{\centering
\includegraphics*[width=0.3\textwidth]{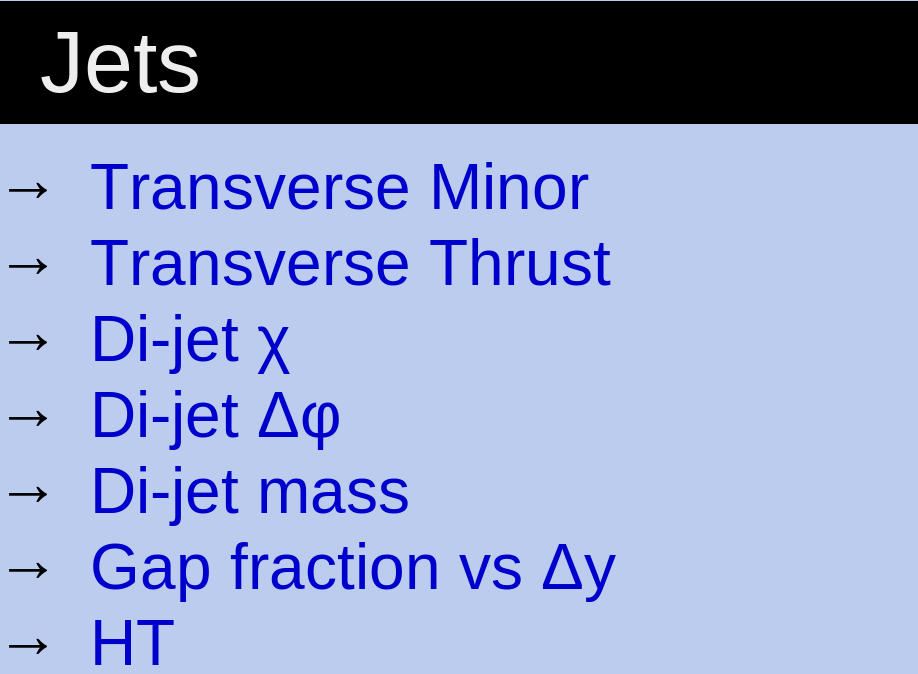}
} }
\subfloat[An expanded dropdown menu\label{fig:obsListExp}] 
{\parbox{0.4\textwidth}{\centering
\includegraphics*[width=0.3\textwidth]{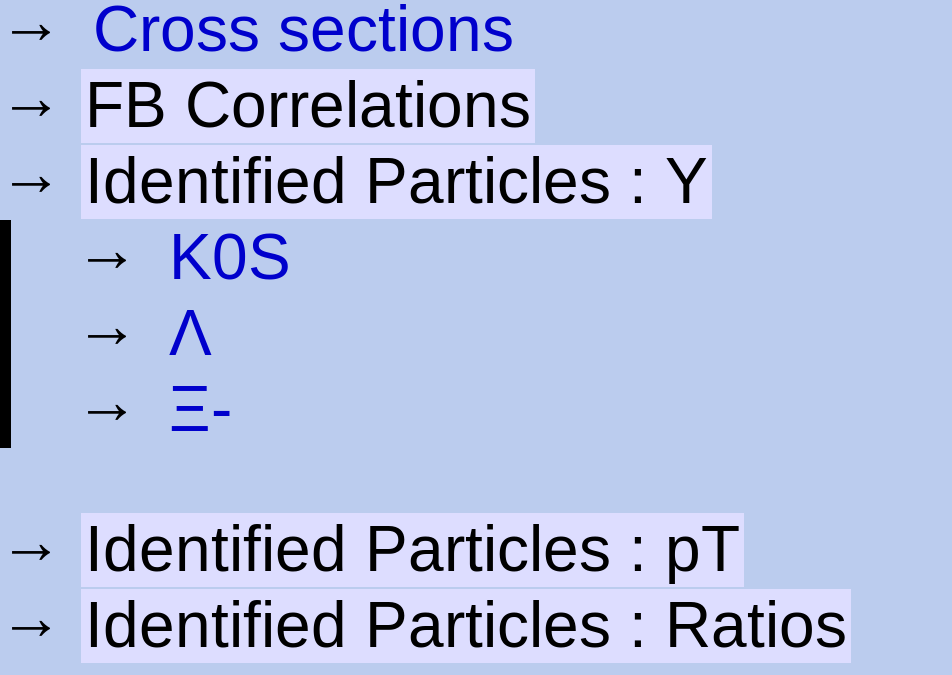}
}} 
\caption{Illustrations of the process and observables list; normal
  view (left) and after 
  clicking on a shaded dropdown menu (right), in this case {\sl
    Identified Particles: Y}.\label{fig:obsList}}
\end{figure}

Clicking on any blue link below one of the process headers  
(e.g.\ below the "Jets" header), \figRef{fig:obsListDef}, 
or any blue link in the shaded drop-down menus,
\figRef{fig:obsListExp}, will open the plot
page for that observable in the right-hand part of the page.

\begin{figure}[p]
\centering
\includegraphics*[width=0.96\textwidth]{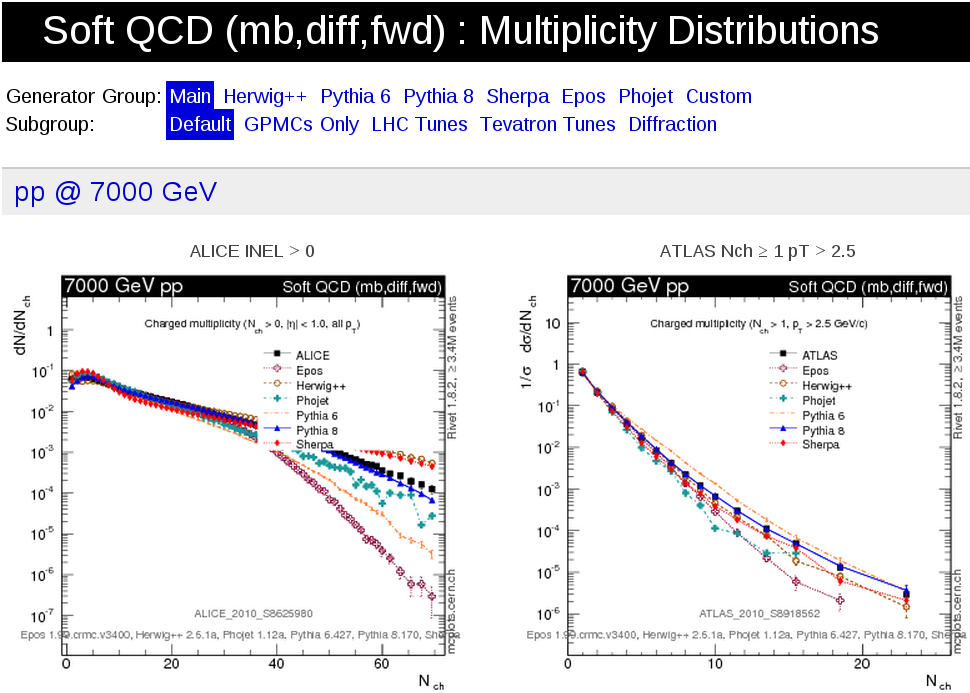}
\caption{The plot page. The generator and tune group selections are
  at the top, followed by the available plots for the chosen
  observable, ordered by CM energy and subordered alphabetically. 
\label{fig:plotPage}}
\end{figure}
\begin{figure}[p]
\centering
\begin{tabular}{p{0.48\textwidth}p{0.48\textwidth}}
\vspace{0mm}%
\subfloat[The Ratio Pane \label{fig:ratioPaneDef}]
{\includegraphics*[width=0.48\textwidth]{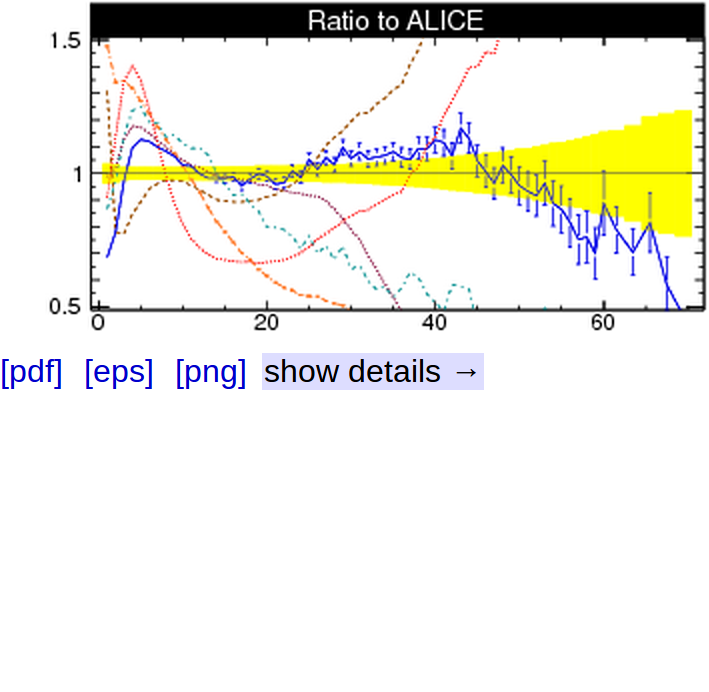}}
&\vspace{0mm}%
\subfloat[After clicking on {\sl show details}\label{fig:showDetails}] 
{\includegraphics*[width=0.48\textwidth]{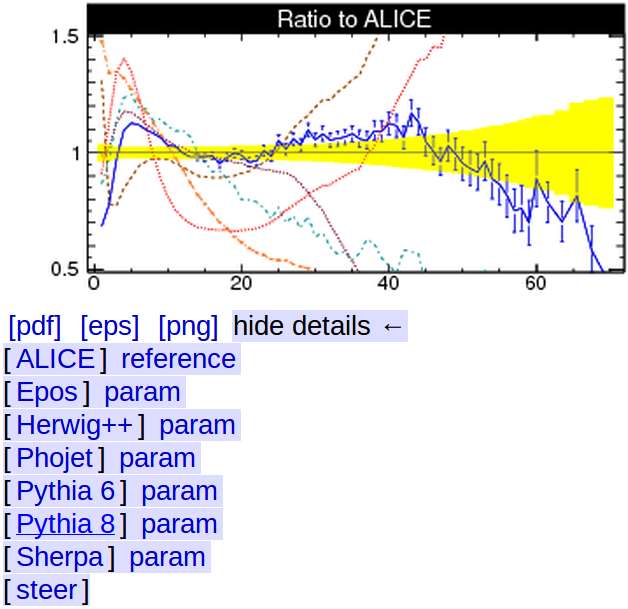}}
\end{tabular}
\caption{The ratio pane, before (a) and after (b)
  clicking on the {\sl show details} link. The example corresponds to 
  the left-hand plot shown in \figRef{fig:plotPage}. 
\label{fig:ratioPane}}
\end{figure}

At the top of the plot page, \figRef{fig:plotPage}, 
you can select which generators and tune
combinations you want to see on the page. By default, you are shown
the results obtained with default settings of the available
generators, but the links for each generator give you access to see
results for different tune and model variations. Use the {\sl Custom}
link to specify your own set of generators and tunes. The 
available plots for the chosen settings are shown starting with the
highest CM energies at the top of the page, and, for each CM energy, 
cascading from left to right alphabetically. 

For many observables, measurements have been made using a variety of
different cuts and triggers. These are indicated both above the plots
and on the plots themselves, so as to minimize the potential for
misinterpretation. In the example of charged-particle multiplicity
distributions shown in \figRef{fig:plotPage}, the two first plots
that appear are thus ALICE (left), for their INEL$>$0
cuts~\cite{Aamodt:2010pp}, 
and ATLAS (right), using their $N_\mrm{ch}\ge 1$ and $p_T>2.5\,\mrm{GeV}$
cuts~\cite{Aad:2010ac}. We explain how to find the correct
references and run cards for each plot and generator below. Note that,
for the Monte Carlo runs, the number of events in the smallest
  sample is shown along the right-hand edge of each plot. I.e., if two
  generators were used, and the statistics were $N_1$ and $N_2$
  events, respectively, the value printed is $\min(N_1,N_2)$. 

Underneath each plot is shown a ratio pane, showing the same results
normalized to the data (or to the first MC curve if there are no data
points on the plot). This is illustrated in \figRef{fig:ratioPane}. 

The vertical range of the ratio plot is fixed to $[0.5,1.5]$, thus larger
deviations than this will exceed the boundaries of the ratio plot. The
central shaded (yellow) band corresponds to the experimental
uncertainty, at the $1\sigma$ level, as reported by \rivet. For
MC-to-MC comparisons, it represents the statistical uncertainty of the 
reference MC. 

Immediately below the ratio pane are links to download the plot (main
plot + ratio pane) in higher resolution and/or in vector graphics
formats. Currently, {\sl pdf}, {\sl eps}, and {\sl png} formats are
available. For publications or presentations, we strongly recommend 
using these high-resolution versions rather than the web-optimized
ones displayed on the page, to avoid undesirable pixelation effects. 

Additional information about the plot, such as data tables,
references, and run cards, can be accessed by clicking on the {\sl show details} dropdown menu, illustrated in
\figRef{fig:showDetails}. In this example, clicking on {\sl [ALICE]}
gives you a text file containing a table of the experimental data
points, along with additional information (from \rivet) about the
plot. Clicking on {\sl reference} sends you to the inSPIRE page for
the experimental paper in which the measurement was presented. 
Clicking on a generator name will give you a text file 
containing a table of the results obtained with that generator,
together with additional technical information from \mcplots 
(including an additional table which is used by \mcplots to combine the results
of several different runs). You can use or ignore the additional
information in these files as you wish. Clicking on {\sl param} gives
you the exact  
generator run card that was used to make the plot, so that you can see
precisely how the results were generated. These cards can also be
useful as examples to start generating your own standalone
results, or to check that you can reproduce ours. 
Finally the {\sl steer} link contains the steering card used
by the \textsc{root}-based tool that makes the actual plots you
see. Though this tool is not yet publicly accessible on the site,
just contact us if you would like a copy, e.g.\ to use it to 
make your own standalone plots outside of \mcplots. 

\subsection{The Generator Validation View \label{sec:validationView}}

Clicking on {\sl Generator Validation} (see the main \mcplots menu in 
\figRef{fig:mainMenu})
opens the validation view. (You can click on {\sl Front Page} to get back to
the default view at any time.) In this view, instead of the {\sl
  Analysis Filter} 
you will see a list of event generators beneath the main \mcplots menu. You
can click on each generator to see a list of the available tunes and
model variations for that generator. This is illustrated in
\figRef{fig:validationView}.

\begin{figure}[t]
\centering
\includegraphics*[width=0.3\textwidth]{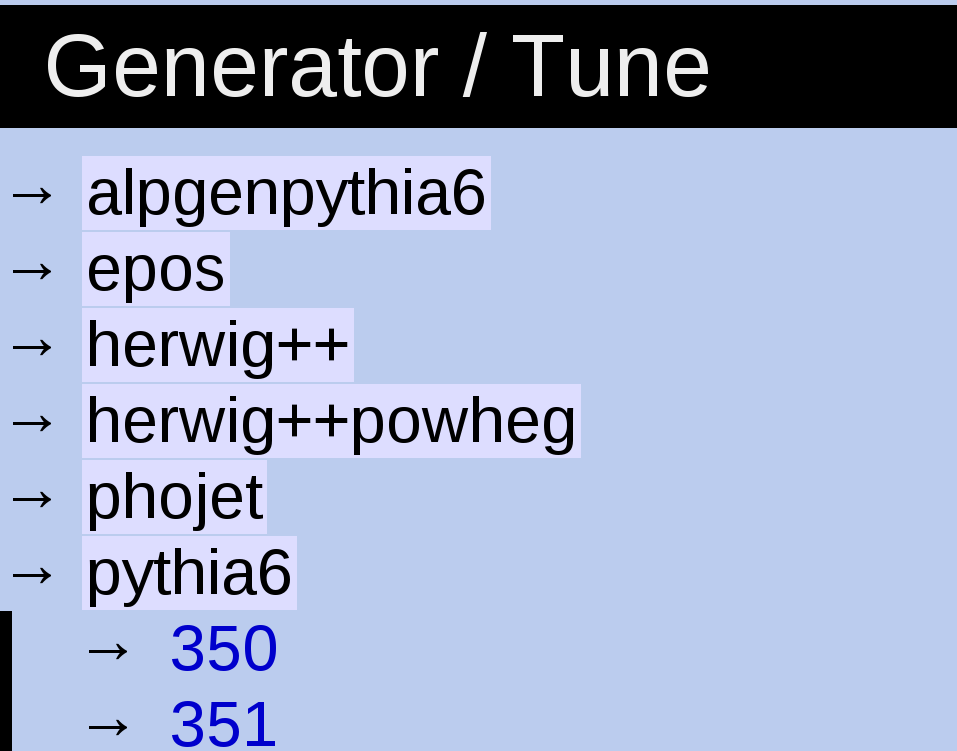} 
\caption{Generator Validation View: excerpt of 
the generator/tune selection menu.\label{fig:validationView}} 
\end{figure}

As an example, clicking on {\sl pythia8 $\to$ default} will show the
validation view for default settings of the \py~8 generator, in the
right-hand side of the page, illustrated in \figRef{fig:py8valid}. In
this view, no plots are shown immediately. Instead you are presented
with a table of $\left<\chi^2\right>$ values, averaged over all
measurements within each process category. Note that we use a slightly
modified definition of $\left<\chi^2\right>$, 
\begin{equation}
\left<\chi^2\right>_\mrm{\mcplots} \ = \ \frac{1}{N_\mrm{bins}}
\sum_{i=1}^{N_\mrm{bins}} \frac{(\mrm{MC}_i -
  \mrm{Data}_i)^2}{\sigma_{\mrm{Data},i}^2 + 
  \sigma_{\mrm{MC},i}^2 + (\epsilon_\mrm{MC} \mrm{MC}_i)^2}~,
\label{eq:chi2}
\end{equation}
where $\sigma_{\mrm{Data},i}$ is the  uncertainty on the
experimental measurement (combined statistical and systematic) of bin
number $i$, and $\sigma_{\mrm{MC},i}$ is the (purely statistical) MC
uncertainty in the same bin. 
The additional relative uncertainty, $\epsilon_\mrm{MC}$, 
associated with the MC prediction, is commented on below. 
From the MC and data histograms alone, it is difficult to determine 
unambiguously whether the number of degrees of freedom for 
a given distribution is $N_\mrm{bins}$ or $(N_\mrm{bins}-1)$, 
hence we currently use 
$1/N_\mrm{bins}$ as the normalization factor for all $\left<\chi^2\right>$
calculations.  

At the top of the page, you can select which versions of the generator
you want to include in the table. Click the {\sl Display} button to
refresh the table after making modifications. 

\begin{figure}[tp]
\centering
\includegraphics*[scale=0.55]{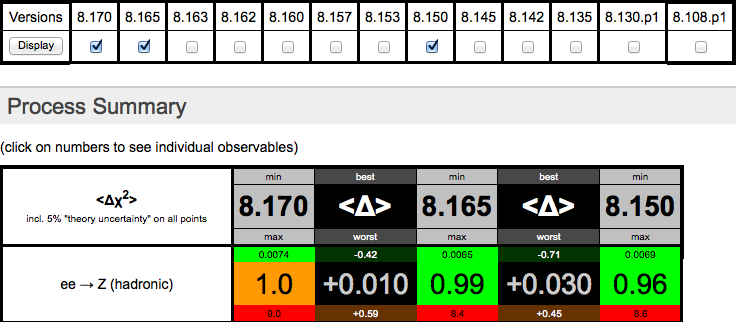}\vskip-1mm
\includegraphics*[scale=0.55]{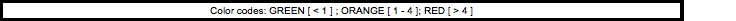} 
\caption{Generator Validation: example showing (an excerpt of) the
  validation view for default settings of the \py~8 generator.\label{fig:py8valid}} 
\end{figure}
Below the line labelled {\sl Process Summary}, we show the main table
of $\left<\chi^2\right>$ values for the versions you have selected, as
well as the relative changes between successive versions, thus
allowing you to look for any significant changes that may have
resulted from improvements in the modelling/tuning (reflected by
decreasing $\chi^2$ values) or mistunings/bugs (reflected by
increasing $\chi^2$ values). The largest and smallest individual
$\left<\chi^2\right>$ values (and changes) in the relevant data set 
are also shown, in smaller fonts, above and below the average values. 
To aid the eye, values smaller than 1 are shaded green (corresponding to less
than $1\sigma$ average deviation from the data), values between 1 and 4
are shaded orange (corresponding to less than $2\sigma$ deviation),
and values greater than 4 are shaded red, following the spirit of the
Les Houches ``tune killing'' exercise reported on 
in~\cite{AlcarazMaestre:2012vp}. In the example shown in
\figRef{fig:py8valid}, the changes are less than a few per cent,
indicative of no significant overall change. 

Bear in mind that the
statistical precision of the MC samples plays a role, hence small
fluctuations in these numbers are to be expected, depending on the
available numbers of generated events for each version. A future
revision that could be considered would be to reinterpret the
statistical MC uncertainties in terms of uncertainties on the
calculated $\left<\chi^2\right>$ values themselves. This would allow a better 
distinction between a truly good description (low
$\left<\chi^2\right>$ with small uncertainty) 
and artificially low $\left<\chi^2\right>$ values caused by low MC
statistics (low $\left<\chi^2\right>$ with large uncertainty). Such
improvements would certainly be mandatory before making any rigorous 
conclusions using this tool, as well as for objective interpretations of direct 
comparisons between different generators. For the time being, the tool is
not intended to provide such quantitative discriminations, but merely to 
aid authors in making a qualitative assessment of where their codes
and tunes work 
well, and where there are problems. 

Note: to make these numbers more physically meaningful, the 
generator predictions are
assigned a flat $\epsilon_\mrm{MC} = 5\%$ ``theory uncertainty''  in
addition to the purely statistical MC uncertainty, see
\eqRef{eq:chi2}, as a baseline sanity limit for the achievable
theoretical accuracy 
with present-day models. A
few clear cases of GIGO\footnote{Garbage In, Garbage Out.} 
are excluded from the $\chi^2$ calculation, but some problematic cases
remain. Thus, e.g., if a calculation returns a too small cross section
for a dimensionful quantity, the corresponding $\chi^2$ value will be
large, even though the shape of the distribution may be well
described. It could be argued how this should be treated, how much
uncertainty should be allowed for each observable, how to compare
consistently across models/tunes with different numbers 
of generated events, whether it is
reasonable to include observables that a given model is not supposed
to describe, etc. These are questions that we do not believe can be
meaningfully (or reliably) addressed by a fully automated site
containing tens of thousands of model/observable combinations. In the
end, the interpretation of the information we display is up to you,
the user. That is also why, at least for the time being, we do not
display any direct comparisons between different MC generators. 

To see the values for all of the individual distributions that enter a
given process type, click on any of the $\left<\chi^2\right>$ values in the
table. Or, to see a comparison between two successive versions, click
on any of the $\Delta\left<\chi^2\right>$ values. The result of doing
the latter is illustrated in \figRef{fig:py8valDetail}. You can now
scroll down the list of observables to get a more detailed view of
which observables actually changed, and how. Also in this view, you
can click on the numbers in the table. 
\begin{figure}[tp]
\centering
\includegraphics*[scale=0.55]{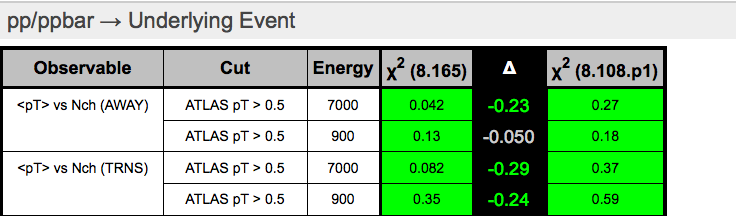}
\caption{Generator Validation: example showing (an excerpt of) the
  detailed observable-by-observable validation view for default
  settings of the \py~8 generator, 
  comparing two different versions.\label{fig:py8valDetail}} 
\end{figure}
Doing so takes you to the final level of validation detail, which is a
plot showing the two generator versions compared on the given
observable. In the example of \figRef{fig:py8valDetail}, clicking on the
value {\sl -0.29} in the third row will produce the plot shown in
\figRef{fig:py8valFinal}. As usual, you can use the links below the
plot to download it in various formats or obtain all the numerical 
information that was used to make it. 
\begin{figure}[tp]
\centering
\includegraphics*[scale=0.44]{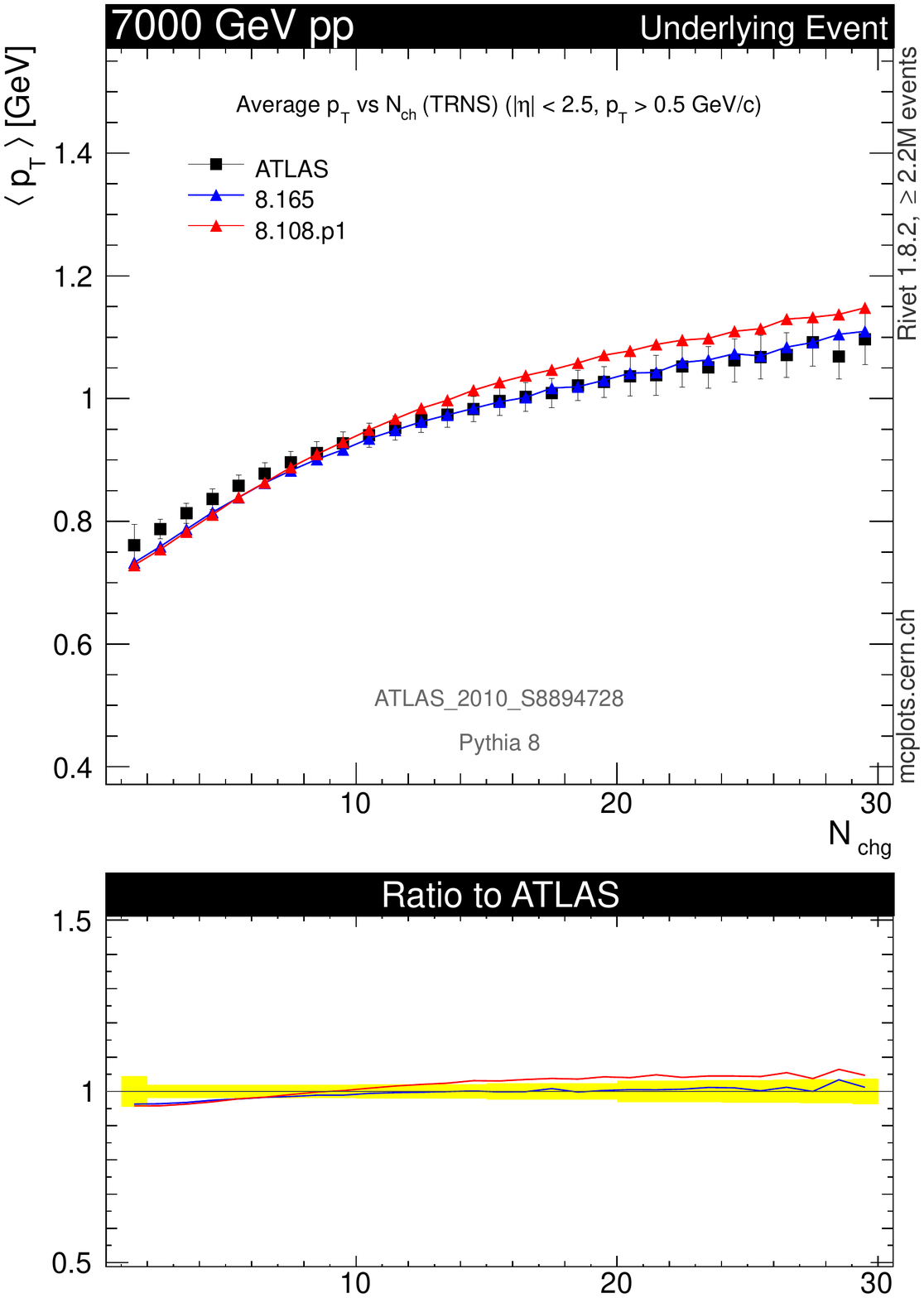}
\caption{Generator Validation: example showing the plot produced by
  clicking on the number {\sl -0.29} in the $\left<\chi^2\right>$
  table shown in \figRef{fig:py8valDetail} (downloaded in PDF
  format). Data from ~\cite{Aad:2010fh}.
  \label{fig:py8valFinal}} 
\end{figure}

\clearpage 
\section{Architecture and Implementation \label{sec:architecture}}

\subsection{Histogram and Data Format \label{sec:hisFormat}}

On \mcplots, a common numerical file format is used for both
experimental data and MC generator results. The format is based on
\rivet's ``flat format" for histograms\footnote{See section 4.1 of the
\rivet manual~\cite{Buckley:2010ar} for details. \mcplots uses
\rivet's {\tt aida2flat} script to convert the \rivet-output to the
flat format, and thus currently relies on \rivet's initial,
\textsc{Aida}-based, histograming. A future update is planned to
upgrade the site to use \rivet~2.0, which employs a new histogram
format called \textsc{Yoda}.} with modifications to store additional 
information which is necessary for histograms merging. The format itself
is a plain text ASCII file which has the advantage of being both human
and machine readable and writable. Each file is divided into sections with
the format

\vskip\vbspc\noindent\begin{minipage}{\textwidth}{\small
\begin{verbatim}
  # BEGIN type
  param1=value1
  param2=value2
  bin1 contents
  bin2 contents
  ...
  # END type
\end{verbatim}
}\end{minipage}\vskip\vbspc

\noindent
The following types of sections have been defined:

\begin{itemize}
\item PLOT - description of plot parameters, for example axis
  labels or scale type (linear/logarithmic).
\item HISTOGRAM - plot data (bin contents), used by the 
  plotting tool to produce the plots.
\item HISTOSTATS, PROFILESTATS - additional histogram (or profile) 
  statistical data, used to merge histograms from different
  subruns.
\item METADATA - description of MC generator parameters, for example
  generator name and version, simulated process and cross section, etc.
\end{itemize}

Section contents can be of two kinds: parameter value definitions
or bin value definitions. Parameter value definitions have the same
format for all types of sections. Bin value definitions contain
several values on each line, e.g.\ describing the bin position, contents, and
uncertainties, with the following specific formats for each section type:

\vskip\vbspc\noindent\begin{minipage}{\textwidth}{\small
\begin{verbatim}
  HISTOGRAM:
   xLo xMid xHi   yValue error- error+

  HISTOSTATS:
   xLo xMid xHi   entries sumW sumW2 sumXW sumX2W

  PROFILESTATS:
   xLo xMid xHi   entries sumW sumW2 sumXW sumX2W sumYW sumY2W sumY2W2
\end{verbatim}
}\end{minipage}\vskip\vbspc

\noindent
where:

\begin{itemize}
\item xLo, xHi - bin edges low and high positions.
\item xMid - bin centre position.
\item yValue - bin value.
\item error-, error+ - bin negative and positive errors.
\item entries - number of bin fills.
\item sum* - sums of various quantities accumulated during bin fills:
  weight, weight*weight, X*weight, X*X*weight, Y*weight, Y*Y*weight,
  Y*Y*weight*weight.
\end{itemize}
As mentioned above, the sum-type quantities provide the additional 
statistical information needed to combine histograms from different runs. 

\subsection{Directory Structure \label{sec:dirs}}

\mcplots is structured as a directory tree, with separate subdirectories
for different purposes (WWW content, MC production scripts, documentation,
\dots). In the following, we assume that this directory structure has been
installed in a global home directory, which we call

\vskip\vbspc\noindent\begin{minipage}{\textwidth}{\small
\begin{verbatim}
  $HOME     = /home/mcplots     # MCPLOTS home directory
\end{verbatim}
} \end{minipage}\vskip\vbspc

\noindent 
The location \code{/home/mcplots} corresponds to the
current implementation on the \textbf{mcplots.cern.ch}
server. There are five important subdirectories in the home directory:

\vskip\vbspc\noindent\begin{minipage}{\textwidth}{\small
\begin{verbatim}
  $DOC      = $HOME/doc         # Documentation and Help
  $PLOTTER  = $HOME/plotter     # Source code for ROOT-based plotting tool
  $POOL     = $HOME/pool        # Output from BOINC cluster
  $RELEASE  = $HOME/release     # Collection of generated (MC) data
  $SCRIPTS  = $HOME/scripts     # Production and update scripts
  $WWW      = $HOME/www         # Front of house: WWW pages and content
\end{verbatim}
}\end{minipage}\vskip\vbspc

The \code{\$DOC} directory contains documentation and help
content that extends and updates this write-up and should be consulted
by future developers.
The \code{\$PLOTTER} directory 
contains the C++ source code and Makefile for the small plotting
utility used to generate the plots on the mcplots web pages. It only depends
on ROOT and takes histogram/data files in the format used on the
mcplots site as input. It can be copied and/or modified for personal
use, if desired, as e.g., in \cite{Cooper:2011gk}. 
Any changes to the code located in the \code{\$PLOTTER}
directory itself, on the \mcplots server, will be reflected in the plots
appearing on the site, after the plotter has been recompiled and the
browser cache is refreshed. 

The \code{\$POOL} directory contains the MC data sets generated by the
Test4Theory \boinc\ cluster.

The \code{\$RELEASE} directory 
contains MC data sets generated manually and combinations
of data sets (so-called releases). For example typically the data
visible on the public site \textbf{mcplots.cern.ch} is a combination
of \boinc\ data sets (which have large statistics) and a number of manually
generated data sets applied on top, for distributions that cannot be run
on the \boinc\ infrastructure, or to
add new versions of generators which were not yet available at the time of
generation of the \boinc\ data set.

The \code{\$SCRIPTS} directory contains various scripts used to
organize and run generator jobs, and to update the contents of the
\code{\$WWW} directory which contains all the \html and PHP
source code, together with style and configuration files, 
for the \textbf{mcplots.cern.ch} front end of the site.

This directory structure is clearly visible in the \mcplots SVN
repository, which is located at \url{https://svn.cern.ch/reps/mcplots/trunk/},
and which can be accessed through a web browser at
\url{https://svnweb.cern.ch/trac/mcplots/browser/trunk/}. Neither the
\code{\$RELEASE} nor the \code{\$POOL} directory are part of the SVN
repository, so as to keep the repository minimal. 

\section{Implementing a new analysis \label{sec:newAnalysis}}

In this section, we describe how to implement additional analyses in
the \mcplots generation machinery so that the results will be
displayed on the \textbf{mcplots.cern.ch} web page. 

The implementation of new analyses relies  
on four main steps, the first of which only concerns comparisons to
experimental data (it is not needed for pure MC
comparisons). Furthermore, both the first and second steps are
becoming standard practice for 
modern experimental analyses, minimizing the remaining burden. The steps
are described in \secsRef{sec:newData} -- \ref{sec:newPlots}. 

\subsection{HEPDATA \label{sec:newData}}
First, the relevant set of experimentally measured data points 
must be provided in the Durham HEPDATA repository,
which both \rivet and \mcplots rely on, see \cite{Buckley:2010jn} for
instructions on this step. Important aspects to consider is whether
the data points and error bars are all physically meaningful (e.g., no
negative values for a positive-definite quantity) and whether
a detector simulation is required to compare MC generators to them or
not.  

If a detector simulation is needed (``detector-level'' data), 
it may not be possible to complete the following steps, unless 
some form of parametrized response- or smearing-function is available,
that can bring MC output into a form that can be compared directly
to the measured data. For this reason, data corrected for detector
effects within a phase-space region corresponding roughly to 
the sensitive acceptance of the apparatus (``particle-level'' with a
specific set of cuts defining the acceptance) is normally preferred\footnote{ 
Corrections to full phase space ($4\pi$ coverage,
without any tracking or calorimeter thresholds) are only useful to the
extent the actual measured acceptance region is reasonably close to
full phase space in the first place, and should be avoided otherwise,
to avoid possible inflation of errors by introducing model-dependent
extrapolations.}. 

For data corrected to the particle level, the
precise definition of the particle level (including definitions of
stable-particle lifetimes, phase-space cuts / thresholds, and any
other corrections applied to the data) must be carefully noted, so
that exactly the same definitions can be applied to the MC output in
the following step.  

\subsection{RIVET}
A \rivet analysis must be provided, that codifies 
the observable definition on MC generator output. As already
mentioned, this is becoming standard practice for an increasing number
of SM analyses at the LHC, a trend we strongly encourage and hope will continue
in the future. \rivet analyses  
already available to \mcplots can e.g.\ be found in the
\code{analyses.html} documentation file of the \rivet installation
used by \mcplots, located in the \rivet installation directory.
For instructions on implementing a new \rivet
analyses, see \cite{Buckley:2010ar}, which also includes ready-made
templates and examples that form convenient starting points. 


For comparisons to experimental data, see the comments in
\secRef{sec:newData} above on ensuring an apples-to-apples 
comparison between data and Monte Carlo output. 

\subsection{Event generation \label{sec:eventGeneration}}

Given a \rivet analysis, 
inclusion into the production machinery and display of \mcplots can
be achieved by editing a small number of steering files. We illustrate
the procedure with the concrete example of an OPAL analysis
of charged-hadron momentum spectra in hadronic $Z$ decays 
\cite{Ackerstaff:1998hz}, for which a \rivet analysis indeed already
exists, called \code{OPAL\_1998\_S3780481}. 

First, input cards for the MC generator runs must be provided, which
specify the hard process and any additional settings pertaining to the desired
runs. (Note that random number seeds are handled automatically by the
  \mcplots machinery and should not be modified by the user.) 
One such card must be provided for each MC generator for which 
one wishes results to be displayed on the site. Several cards are
already available in the following directory, 

\vskip\vbspc\noindent\begin{minipage}{\textwidth}{\small
\begin{verbatim}
  CARDS = $SCRIPTS/mcprod/configuration
\end{verbatim}
}\end{minipage}\vskip\vbspc

\noindent
where the location of the \code{\$SCRIPTS} directory was defined in
\secRef{sec:dirs}. 
For the example of the OPAL analysis, we wish to run a standard set of
hadronic $Z$ decays, and hence we may reuse the existing cards:

\vskip\vbspc\noindent\begin{minipage}{\textwidth}{\small
\begin{verbatim}
  $CARDS/herwig++-zhad.params
  $CARDS/pythia6-zhad.params
  $CARDS/pythia8-zhad.params
  $CARDS/sherpa-zhad.params
  $CARDS/vincia-zhad.params
\end{verbatim}}
\end{minipage}\vskip\vbspc

The generator and process names refer to the internal names used for
each generator and process on the \mcplots site. (The site is
constructed such that any new process names automatically appear on
the web menus, and each can be given a separate \html and \latex name,
as described in \secRef{sec:newPlots}.) In this example, 
hadronic $Z$ decays are labelled \code{zhad}. For processes
for which the \code{\$CARDS} directory does not already 
contain a useful set of card files, new ones must be defined, by referring to
the documentation of examples of each generator code separately. This
is the only generator-dependent part of the \mcplots machinery. 

Having decided which cards to use for the hard process(es),
information on which observables are are available in the \rivet
analysis corresponding to our OPAL measurement 
are contained in the file 

\vskip\vbspc\noindent\begin{minipage}{\textwidth}{\small
\begin{verbatim}
  $RIVET/share/Rivet/OPAL_1998_S3780481.plot
\end{verbatim}
} \end{minipage}\vskip\vbspc

\noindent
Decide which individual distributions of that analysis 
you want to add to \mcplots, 
and what observables and cuts they correspond to. In our
case, we shall want to add the distributions \code{d01-x01-y01},
\code{d02-x01-y01}, \code{d05-x01-y01} and \code{d06-x01-y01}, which represent
the $x$ distributions (with $x$ defined as momentum divided by half the
centre-of-mass energy) in light-flavour events (\code{d01}), 
the $x$ distributions in $c$-tagged events (\code{d02}), 
and the $\ln(x)$ distributions in light-flavour (\code{d05}) and
$c$-tagged (\code{d06}) events, respectively.

The new analysis is added to the \mcplots production 
machinery by adding one line to the file

\vskip\vbspc\noindent\begin{minipage}{\textwidth}{\small
\begin{verbatim}
  $CARDS/rivet-histograms.map
\end{verbatim}
}\end{minipage}\vskip\vbspc

\noindent 
for each new observable. In the case of the OPAL analysis, we would add the 
following lines


\vskip\vbspc\noindent\begin{minipage}{\textwidth}{\scriptsize
\begin{verbatim}
#== MC Initialization and Cuts in GeV ===  ========= Rivet ========  ====== mcplots =======
# beam process  Ecm  pTMin,Max,mMin,Max    RivetAnalysis_Hist          Obs              Cut

  ee    zhad    91.2        -        OPAL_1998_S3780481_d01-x01-y01    x     opal-1998-uds
  ee    zhad    91.2        -        OPAL_1998_S3780481_d02-x01-y01    x     opal-1998-c
  ee    zhad    91.2        -        OPAL_1998_S3780481_d05-x01-y01    xln   opal-1998-uds
  ee    zhad    91.2        -        OPAL_1998_S3780481_d06-x01-y01    xln   opal-1998-c
\end{verbatim}
}\end{minipage}\vskip\vbspc

\noindent
If several different analyses include the same observable (or approximately the
same, e.g., with different cuts), we recommend to assign them the same
consistent name in the \code{Obs} column (such as \code{x} and 
\code{lnx} above). This will cause the corresponding plots to be
displayed on one and the same page on the \mcplots web pages, rather
than on separate ones, with \code{Cut} giving a further labelling of 
the individual distributions on each page.  

As an option to optimize the MC production, {\tt rivet-histograms.map} allows
the specification of a set of phase-space cuts within which to restrict the
hard-process kinematics. These optional settings can be provided by changing
the optional \code{pTMin}, \dots, \code{mMax} columns above (in our 
example, such cuts are not desired, which is indicated by 
the \code{-} symbol). Note that such cuts must be applied with extreme care, 
since they refer to the hard partonic subprocess, not the final physical 
final state, and hence there is always the risk that bremsstrahlung or other
corrections (e.g., underlying event) can cause events to migrate across
cut thresholds. In the end, only the speed with which the results are obtained
should depend on these cuts, not the final physical distributions
themselves (any such dependence is a sign that looser cuts are
required). We would like to point out that a separate generator run is 
required for each set of MC cuts. Hence, it is useful to choose as small a 
set of different cuts as possible. The following is an excerpt from
\code{rivet-histograms.map} which concerns a CDF 
analysis of differential jet shapes \cite{Acosta:2005ix}, in which two different 
generator-level $\hat{p}_\perp$ cuts
are invoked to ensure adequate population of a much larger number of jet
$p_\perp$ bins (here ranging from 37 to 112 GeV):

\vskip\vbspc\noindent\begin{minipage}{\textwidth}{\footnotesize
\begin{verbatim}
#=== MC Initialization and Cuts in GeV ==    ======= Rivet ======  === mcplots ===
# beam process    Ecm  pTMin,Max,mMin,Max      RivetAnalysis_Hist     Obs      Cut
 ppbar    jets   1960        17     CDF_2005_S6217184_d01-x01-y01 js_diff cdf3-037
 ppbar    jets   1960        17     CDF_2005_S6217184_d01-x01-y02 js_diff cdf3-045
 ppbar    jets   1960        37     CDF_2005_S6217184_d02-x01-y02 js_diff cdf3-073
 ppbar    jets   1960        37     CDF_2005_S6217184_d02-x01-y03 js_diff cdf3-084
\end{verbatim}
}\end{minipage}\vskip\vbspc

After changing \code{rivet-histograms.map}, it is necessary to run the 
available MC generators in order to produce the new histograms, and then 
update the \mcplots database in order to display the results. We will 
describe how to include new generator tunes, run the generators and update 
the database in \secRef{sec:newVersion}. For now, we will assume that
the results of MC runs are already available, and continue by discussing how
to translate the language of \code{rivet-histograms.map} into the labels
that are displayed on the {\bf mcplots.cern.ch} pages.

\subsection{Displaying the results on MCPLOTS \label{sec:newPlots}}

\begin{figure}[t]
\centering
  \includegraphics[width=0.9\textwidth]{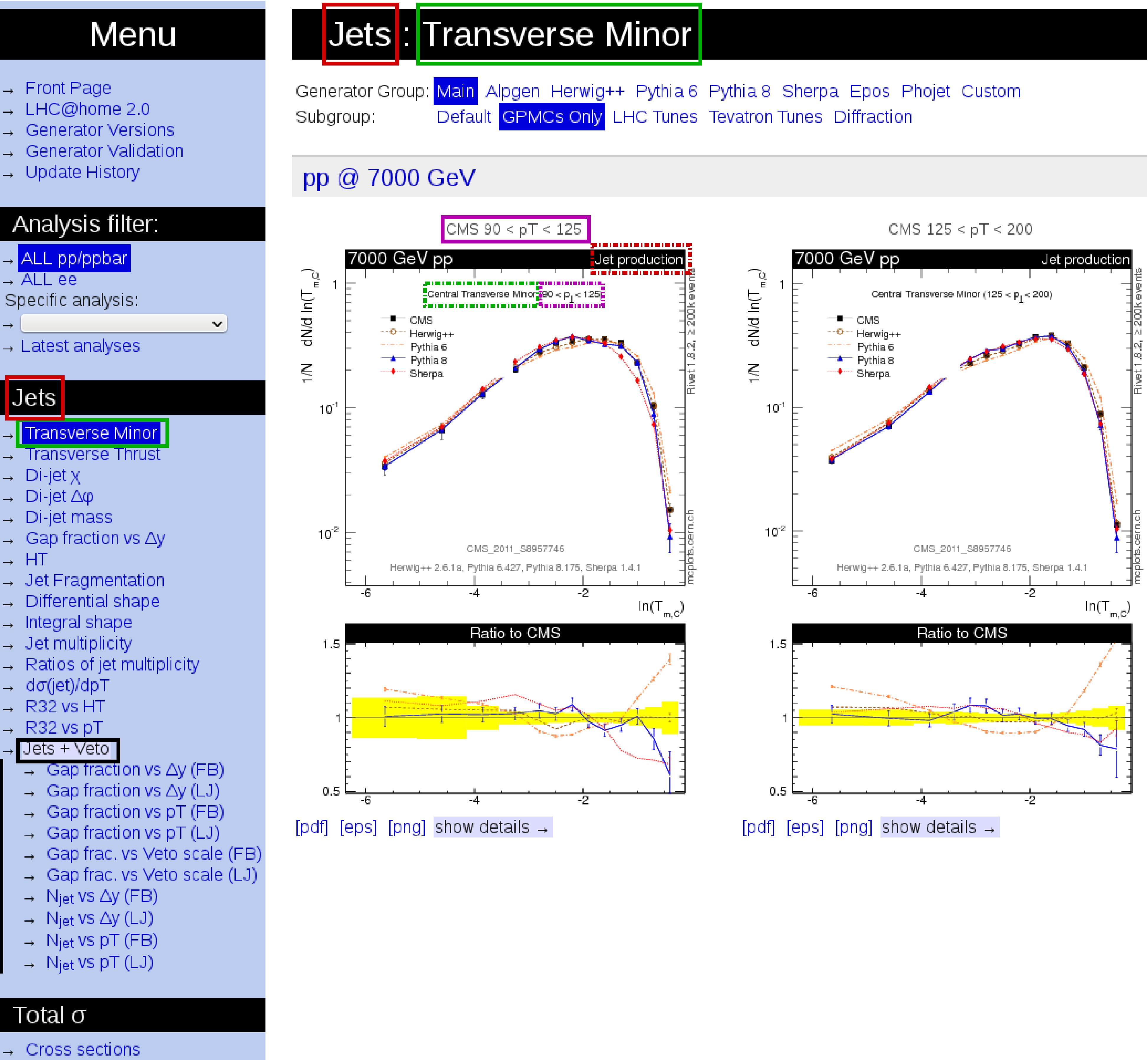}
\caption{\label{Fig:mcplots-shot} Snapshot of an \mcplots page, to serve as 
an illustration of web page labels. Most label items are produced by 
processing the file {\tt \$WWW/mcplots.conf}. \SecRef{sec:newPlots} 
discusses changes of {\tt \$WWW/mcplots.conf}, using the text encased by 
coloured boxes as examples.}
\end{figure}

\noindent
The next step after implementing a new analysis is to define a correspondence 
between the internal names (as declared in the {\tt rivet-histograms.map} file)
and the names that will be displayed on the web page and on plots. 
Correspondence definitions are collected in the configuration file 

\vskip\vbspc\noindent\begin{minipage}{\textwidth}{\small 
\begin{verbatim}
  $WWW/mcplots.conf
\end{verbatim}
}\end{minipage}\vskip\vbspc
\noindent
All internal process-, observable- and cut names defined by adding new
lines to the histogram map should be defined in the configuration file. 
Process names are translated by adding a line 

\vskip\vbspc\noindent\begin{minipage}{\textwidth}{\small 
\begin{verbatim}
  process_name  =  !  HTML name  !  plot label in LaTeX format
\end{verbatim}
}\end{minipage}\vskip\vbspc
\noindent
to the list of processes at the beginning of {\tt mcplots.conf}. Please note 
that the sequence of process name definitions in this file also determines
the order of processes in the web page menu. In \figRef{Fig:mcplots-shot},
the ``Jets" label appears before the ``Total $\sigma$" label because the 
relevant definitions are in consecutive lines in {\tt mcplots.conf}.
The labels associated with the ``jets" process (encased by red 
boxes in \figRef{Fig:mcplots-shot}) have been generated by including the 
line

\vskip\vbspc\noindent\begin{minipage}{\textwidth}{\small 
\begin{verbatim}
  jets  =  !  Jets  !  Jet production
\end{verbatim}
}\end{minipage}\vskip\vbspc
\noindent
The internal name ({\sl jets}) is left of the equality sign, the \html 
name ({\sl Jets}, i.e the text in solid red boxes) is defined in 
the centre, while the plot label ({\sl Jet production}, i.e. the text in
dashed red boxes) stands to the right. \mcplots allows for both an \html name
and a plot label so that (a) the web page is adequately labelled, and (b) 
the information on the plot is sufficient to distinguish it from
all other figures of the reproduced article, even if the plot is stored
separately. Since the plotting tool uses \latex, the
plot label should be specified in \latex format, except that the hash 
symbol should be used instead of the back-slash symbol. Note that the 
exclamation marks are mandatory, as they are also used to separate different 
types of observable names.

When defining observable names, it is possible to group a set of observables
into a submenu by specifying an optional submenu name after the equality 
sign, but before the first exclamation mark. This means that the translation 
of internal observable names proceeds by adding lines like 

\vskip\vbspc\noindent\begin{minipage}{\textwidth}{\small 
\begin{verbatim}
observable_name = (HTML submenu) ! HTML name ! plot label in LaTeX format 
\end{verbatim}
}\end{minipage}\vskip\vbspc\noindent 
to {\tt mcplots.conf}. It is not necessary to follow a predefined ordering
when adding observable name correspondences. \FigRef{Fig:mcplots-shot}
illustrates how parts of the declaration lines

\vskip\vbspc\noindent\begin{minipage}{\textwidth}{\scriptsize
\begin{verbatim}
ctm            =             ! Transverse Minor              ! Central Transverse Minor
gapfr-vs-dy-fb = Jets + Veto ! Gap fraction vs &Delta;y (FB) ! Gap fraction vs #Deltay (FB)
\end{verbatim}
}\end{minipage}\vskip\vbspc
\noindent 
are related to the web page layout. The \html submenu name (e.g.\ {\sl
  Jets + Veto}, i.e.\ the text in the solid black box) will appear as
a link on a grey field.  
The \html name of the displayed plot ({\sl Transverse Minor}, i.e.\ the text in
solid green boxes) enters both in the menu and on top of the page, 
while in the \latex observable name ({\sl Central Transverse Minor}, i.e.\ the 
text in the dashed green box) is printed directly onto the plot.

Finally, cut names have to be translated as well. This is achieved by 
expanding {\tt mcplots.conf} with cut declaration lines in the format

\vskip\vbspc\noindent\begin{minipage}{\textwidth}{\small 
\begin{verbatim}
  cut_name    = ! HTML name            ! plot label in LaTeX format 
\end{verbatim}
}\end{minipage}\vskip\vbspc\noindent 
The features resulting from adding the line

\vskip\vbspc\noindent\begin{minipage}{\textwidth}{\small 
\begin{verbatim}
  cms1-pt090  = ! CMS  90 < pT < 125   ! 90 < p_{#perp}  < 125
\end{verbatim}
}\end{minipage}\vskip\vbspc
\noindent 
are shown, encased in magenta boxes, in \figRef{Fig:mcplots-shot}. Again,
the plot label ($90 < p_{\perp}  < 125$, i.e. the text in the dashed 
magenta box) is included in the plot -- in parentheses, and after the 
observable label.  

After translating the internal process, observable and cut names, the display
on \mcplots is fixed. For the sample OPAL analysis discussed in section 
\ref{sec:eventGeneration}, the current layout would be obtained by adding the 
lines

\vskip\vbspc\noindent\begin{minipage}{\textwidth}{\small 
\begin{verbatim}
# Process names
  zhad            = ! Z(hadronic)             ! Z(Hadronic)
# Observable names 
  x               = ! Scaled momentum         ! Scaled momentum
  xln             = ! Log of scaled momentum  ! Log of scaled momentum
# Cut names 
  opal-1998-uds   = ! OPAL u,d,s events       !  OPAL u,d,s events
  opal-1998-c     = ! OPAL c events           !  OPAL c events 
\end{verbatim}
}\end{minipage}\vskip\vbspc
\noindent 
to the configuration file \code{\$WWW/mcplots.conf}.
These steps include a new analysis into the \mcplots display framework, so 
that new plots will be visible after the next update of the database. Before
discussing how to update the \mcplots framework to produce MC runs 
for the new analysis and display the corresponding plots 
(section \ref{sec:newVersion}), we
will discuss some advanced display possibilities that are steered by
{\tt mcplots.conf}.

\subsubsection{Tune groups \label{sec:tuneGroups}}

\mcplots further allows to manipulate which of the available generator runs
should be displayed together in one plot. This is possible by defining
\emph{tune groups} in {\tt mcplots.conf}. Tune groups apply globally to all
processes and observables. All available groups will be displayed as
``Generator groups" or ``Subgroups" between the black observable label bar
and the grey collider information bar. The definition of a tune group requires
three steps. To begin with, a correspondence between the internal generator
name\footnote{How to include add new generators and tunes to the event
generation machinery will be explained in section \ref{sec:newVersion}.} and
the public name has to be defined by including a line in the format

\vskip\vbspc\noindent\begin{minipage}{\textwidth}{\small 
\begin{verbatim}
  generator_name   = ! generator name !
\end{verbatim}
}\end{minipage}\vskip\vbspc
\noindent
in the same section of {\tt mcplots.conf} as other generator names. This should
be followed by the definition of the tune name through a line like
 
\vskip\vbspc\noindent\begin{minipage}{\textwidth}{\small 
\begin{verbatim}
  generator_name.tune_name  = ! tune name !
\end{verbatim}
}\end{minipage}\vskip\vbspc
\noindent 
in the tune name section of the configuration file. Furthermore, a line style
for this particular tune has to be defined in the format


\vskip\vbspc\noindent\begin{minipage}{\textwidth}{\small
\begin{verbatim}
  generator_name.tune_name = ColR ColG ColB lineStyle lineWidth \ 
                                                   markerStyle markerSize
\end{verbatim}
}\end{minipage}\vskip\vbspc
\noindent
where we have used the ``$\backslash$" character to imply line
continuation and 
the available options for the style settings are documented in the ROOT web 
documentation\footnote{See \url{http://root.cern.ch}, specifically the
  pages on
  \href{http://root.cern.ch/root/html/TAttLine.html}{lineStyle}
 and
 \href{http://root.cern.ch/root/html/TAttMarker.html}{markerStyle}. A
 nice helping tool to create your own colour schemes is \url{http://colorschemedesigner.com}.}. 
Once multiple generator tunes have been named, tune subgroups 
can be defined. For this, add lines in the format 

\vskip\vbspc\noindent\begin{minipage}{\textwidth}{\small
\begin{verbatim}
  generator_group_name.subgroup_name  = tune_1, tune_2, tune_3
\end{verbatim}
}\end{minipage}\vskip\vbspc
\noindent
at the tune group section of {\tt mcplots.conf}. As an example, let us look at
the ``$\hwpp$ vs. $\sh$" tune group, which is part of the ``$\hwpp$" generator 
group menu. The necessary definitions to construct this tune group are 

\vskip\vbspc\noindent\begin{minipage}{\textwidth}{\small 
\begin{verbatim}
# Tune names
  herwig++.default   = ! Herwig++  !
  sherpa.default     = ! Sherpa !
# Generator names
  herwig++           = ! Herwig++  !
  sherpa             = ! Sherpa !
# Tune line styles
  herwig++.default   = 0.6  0.3  0.0   2 1.5   24 1.25
  sherpa.default     = 1.0  0.0  0.0   3 1.5   33 1.4
# Tune group
  Herwig++.Herwig++ vs Sherpa  = herwig++.default, sherpa.default
\end{verbatim}
}\end{minipage}\vskip\vbspc
\noindent
This concludes the discussion of manipulations of the configuration file 
{\tt mcplots.conf}.  

\section{Updating the MCPLOTS site\label{sec:newVersion}}

The previous section described how to add 
new processes and new analyses to the \mcplots framework, and how to
modify their organization and labelling on the web site. After a new
analysis has been implemented, or when updating the site with new
tunes, generators,  
or \rivet versions, the next step is to update the database with new MC 
generator runs. 

In this section, we will briefly discuss how to update 
existing MC generators by including new generator versions and tunes. 
(How to implement a completely new event generator is described separately, in
\secRef{sec:newGenerator}). 
This is followed by a description of how to manually produce MC results, 
and how to update the database of the development page {\bf
  mcplots-dev.cern.ch} (which is publicly visible but updates can only
be done by \mcplots\ authors), 
which serves as a pre-release testing server for \mcplots. We finish
by explaining how to make a public \mcplots release, transferring the contents
of the development page to the public one (again an operation
restricted to \mcplots\ authors). Aside from \mcplots\ authors, these
instructions may be useful in the context of standalone (private) 
clone(s) of the \mcplots\ structure, created e.g.\ via the
public SVN repository, cf.\ \secRef{sec:dirs}. 

\subsection{Updating existing generators \label{sec:updateGenerator}}

\mcplots takes generator codes from the GENSER repository, which can be found
in 

\vskip\vbspc\noindent\begin{minipage}{\textwidth}{\small
\begin{verbatim}
  /afs/cern.ch/sw/lcg/external/MCGenerators*
\end{verbatim}
}\end{minipage}\vskip\vbspc
\noindent
Only generator versions in this repository can be added to the \mcplots event
generation machinery. To introduce a new generator version on \mcplots, 
changes of the file

\vskip\vbspc\noindent\begin{minipage}{\textwidth}{\small
\begin{verbatim}
  $SCRIPTS/mcprod/runAll.sh
\end{verbatim}
}\end{minipage}\vskip\vbspc
\noindent
are required\footnote{For generator chains like \alp+\hwpp, changing 
the file {\tt \$SCRIPTS/mcprod/runRivet.sh} might also be necessary to add 
accepted chains of versions. This will be explained below.}. Specifically, it
is necessary to update the {\tt list\_runs()} function. To include, for 
example, version 2.7.0 of \hwpp, with a tune called ``default", the line

\vskip\vbspc\noindent\begin{minipage}{\textwidth}{\small
\begin{verbatim}
  echo "$mode $conf - herwig++ 2.7.0 default $nevt $seed"
\end{verbatim}
}\end{minipage}\vskip\vbspc
\noindent
has to be added. Please note that the {\tt list\_runs()} function of 
{\tt runAll.sh} groups the runs of generators into blocks (e.g. all \hwpp 
runs follow after the comment {\tt \# Herwig++}). This order should be 
maintained. The necessary changes of {\tt list\_runs()} are slightly 
different depending on if we want to include a completely new generator 
version or simply a new tune for an existing generator version. The above
line is appropriate for the former. For the latter, let us imagine we want to
add \hwpp v. 2.7.0, with two tunes called ``default" and ``myTune". Then, we 
should add the string

\vskip\vbspc\noindent\begin{minipage}{\textwidth}{\small
\begin{verbatim}
  mul "$mode $conf - herwig++ 2.7.0 @ $nevt $seed" "default myTune"
\end{verbatim}
}\end{minipage}\vskip\vbspc
\noindent
to the \hwpp block of {\tt runAll.sh}.

After this, it is necessary to include the novel generator versions and tunes 
in the file

\vskip\vbspc\noindent\begin{minipage}{\textwidth}{\small
\begin{verbatim}
  $CARDS/generator_name-tunes.map
\end{verbatim}
}\end{minipage}\vskip\vbspc
\noindent
where ``generator\_name" is the name of the MC generator to be updated. For our
second example, the addition of the lines

\vskip\vbspc\noindent\begin{minipage}{\textwidth}{\small
\begin{verbatim}
  2.7.0  default
  2.7.0  myTune
\end{verbatim}
}\end{minipage}\vskip\vbspc
\noindent
to {\tt \$CARDS/herwig++-tunes.map} is necessary. Depending on how tunes are
implemented in the MC generator, it might also be necessary to include a new
file with the tune parameters into the {\tt \$CARDS} directory. Currently, 
this is the case for \hwpp tunes and non-supported tune variations in \py8.
Say the ``myTune" tune of our example would only differ from default \hwpp
by having unit probability for colour reconnections. The corresponding 
\hwpp input setting

\vskip\vbspc\noindent\begin{minipage}{\textwidth}{\small
\begin{verbatim}
  set /Herwig/Hadronization/ColourReconnector:ReconnectionProbability 1.00
\end{verbatim}
}\end{minipage}\vskip\vbspc
\noindent
should then be stored in a file called {\tt herwig++-myTune.tune} in 
the {\tt \$CARDS} directory. After following the above steps, we have 
included a new MC generator version (and/or new tunes) into the \mcplots event 
generation machinery. 

See \secRef{sec:tuneGroups} for how to include new tunes in new or
existing ``tune groups'' on the site, including how to assign
tune-specific default marker symbols, line styles, etc.

\subsection{Producing MC results \label{sec:mcGeneration}}

Once new processes, analyses, versions or tunes have been added, we need to 
produce results for these new settings. This can be done through volunteer 
computing or manually. We will here, since manual small scale production can 
be very useful for debugging purposes, describe how to produce MC results 
manually. The top-level script for producing and analysing of MC generator 
results is 
\vskip\vbspc\noindent\begin{minipage}{\textwidth}{\small
\begin{verbatim}
  $SCRIPTS/mcprod/runAll.sh [mode] [nevt] {filter} {queue}
\end{verbatim}
}\end{minipage}\vskip\vbspc
\noindent where the two first arguments are mandatory and the two latter ones
are optional, with the following meanings:
\begin{myitemize}
\item{\tt mode}: Parameter governing the distribution of MC generator runs. 
      If {\tt mode} is set to \emph{list}, the script simple returns a list of all
      possible generator runs. \emph{dryrun} will set up the event generation
      machinery, but not execute the generation. \emph{local} will queue all
      desired runs on the current desktop, while \emph{lxbatch} distributes
      jobs on the lxplus cluster at CERN.
\item{\tt nevt}: The number of MC events per run
\item{\tt filter}: Optional filter of the MC runs. For instance, if
  {\tt filter} = \emph{herwig++}, 
{\tt runAll.sh} will only produce \hwpp results.
\item{\tt queue}: Queue to be used on the lxplus cluster.
\end{myitemize}
The {\tt runAll.sh} script executes the event generation and analysis 
steering script
\vskip\vbspc\noindent\begin{minipage}{\textwidth}{\small 
\begin{verbatim}
  $SCRIPTS/mcprod/runRivet.sh
\end{verbatim} 
}\end{minipage}\vskip\vbspc\noindent
for all the desired input settings. It is in principle also possible to run
this script separately.

\subsection{Updating the MCPLOTS database \label{sec:updateDatabase}}

To display the results of new MC runs, it is necessary to update the database
with the new histograms. \mcplots has both a development area (with plots
shown on {\bf mcplots-dev.cern.ch}) and an official release area (with
plots shown on {\bf mcplots.cern.ch}).

The development pages are updated as follows.
Let us assume that we have produced new 
\textsc{Epos} results\footnote{For example by running {\tt 
\$SCRIPTS/mcprod/runAll.sh lxbatch 100k " epos "}.}, and these new results
are stored in {\tt \~{}/myResults}. Then, log into
{\tt mcplots-dev} by
\vskip\vbspc\noindent\begin{minipage}{\textwidth}{\footnotesize 
\begin{verbatim}
  $ ssh myUserName@lxplus.cern.ch
  $ ssh mcplots-dev
\end{verbatim} 
}\end{minipage}\vskip\vbspc\noindent
Then copy the new information and to the directory {\tt \$HOME/release}
on this machine. It is encouraged to include the mcplots revision number 
(e.g. 2000) into the directory name. For example
\vskip\vbspc\noindent\begin{minipage}{\textwidth}{\footnotesize 
\begin{verbatim}
  $ mkdir $HOME/release/2000.epos
  $ cd $HOME/release/2000.epos
  $ cp ~/myResults/info.txt .
  $ find ~/myResults/*.tgz | xargs -t -L 1 tar zxf
\end{verbatim} 
}\end{minipage}\vskip\vbspc
\noindent
Now, the new results are stored in the directory {\tt 
\$HOME/release/2000.epos/dat}. To display these (and only these) 
new histograms on {\bf mcplots-dev.cern.ch}, re-point and update the database 
by
\vskip\vbspc\noindent\begin{minipage}{\textwidth}{\footnotesize 
\begin{verbatim}
   $ cd $WWW
   $ ln -sf $HOME/release/2000.epos/dat
   $ $SCRIPTS/updatedb.sh
\end{verbatim} 
}\end{minipage}\vskip\vbspc
\noindent
These actions update the database, and the new plots will be 
visible on the {\bf mcplots-dev.cern.ch} pages. The last step is to
update the HTML file {\tt \$WWW/news.html} including any relevant new
information similarly to what has been done for previous releases. 
Updating the development web pages is a fairly common task while working on 
mcplots. 

Updating the official web pages ({\bf mcplots.cern.ch}) should of
course only be done with great caution. 
For completeness, we here briefly describe the
procedure. The contents of a specific revision of 
the development web page can be transferred
to the official site by executing a simple script.
For this, log into {\tt mcplots-dev.cern.ch} and execute
\vskip\vbspc\noindent\begin{minipage}{\textwidth}{\footnotesize 
\begin{verbatim}
  $ $SCRIPTS/updateServer.sh -r revision -d /path/to/dir/dat
\end{verbatim}
}\end{minipage}\vskip\vbspc
\noindent
where {\tt\small revision} is the SVN revision number of the mcplots
code that you  
want to update the server to, and {\tt\small /path/to/dir/dat} is the
full path to 
the plot data that you would like to display on the web page. Either
of the arguments can be omitted, and at least one argument is
necessary. For backward  
compatibility, we advise to display plot data that has been stored on
mcplots-dev 
according to the suggestions of the development web page update. To give some 
examples, the code of the public web page can be updated to revision number 
2000 by
\vskip\vbspc\noindent\begin{minipage}{\textwidth}{\footnotesize 
\begin{verbatim}
  $ $SCRIPTS/updateServer.sh -r 2000
\end{verbatim}
}\end{minipage}\vskip\vbspc
\noindent
Further changing the web page to only show results of the above
\textsc{Epos} runs means running
\vskip\vbspc\noindent\begin{minipage}{\textwidth}{\footnotesize 
\begin{verbatim}
  $ $SCRIPTS/updateServer.sh -r 2000 -d $HOME/release/2000.epos/dat
\end{verbatim}
}\end{minipage}\vskip\vbspc
\noindent
After these steps, the official release of {\bf mcplots.cern.ch} is publicly
available. Since official releases are usually scheduled on a monthly
(or longer) time scale, we advocate caution when running the update
script. Always thoroughly check the site and functionality after
performing an update, and roll back to a previous version if any
problems are encountered. There is nothing more damaging to a public
web service than faulty operation.

\section{Implementing a new event generator \label{sec:newGenerator}}

In this section, we provide guidelines for adding a new generator to
the \mcplots framework. An up-to-date set of instructions is maintained
in the \mcplots documentation files\footnote{\url{http://svnweb.cern.ch/world/wsvn/mcplots/trunk/doc/readme.txt}~,
  section on ``Adding a new generator''.}.  
The guidelines are accompanied by
concrete examples and comments, drawn from the experience obtained by
the implementation of the \alp generator in the framework.  

When adding a new generator, one should note that the \mcplots framework relies on the libraries with the code needed
for the event generation being available in the format and with
configurations as used by the GENSER project. The generator libraries
are accessed from the CERN-based AFS location accordingly\footnote{contact
  {\tt genser-dev@cern.ch} for new generator support}:

\vskip\vbspc\noindent\begin{minipage}{\textwidth}{\footnotesize 
\begin{verbatim}
   /afs/cern.ch/sw/lcg/external/MCGenerators*
\end{verbatim} 
}\end{minipage}\vskip\vbspc

\noindent
or from the CVMFS replica which is used by CernVM virtual machines
running in the \boinc\ cluster:

\vskip\vbspc\noindent\begin{minipage}{\textwidth}{\footnotesize 
\begin{verbatim}
   /cvmfs/sft.cern.ch/lcg/external/MCGenerators*
\end{verbatim} 
}\end{minipage}\vskip\vbspc

\noindent
Runs of \mcplots scripts with a generator not supported by GENSER is currently not implemented. For generator not supported by GENSER user can only run the \mcplots scripts by checking them out, making private modifications and private generation runs.

\begin{figure}[t!p]
\centering
  \includegraphics[width=0.90\textwidth]{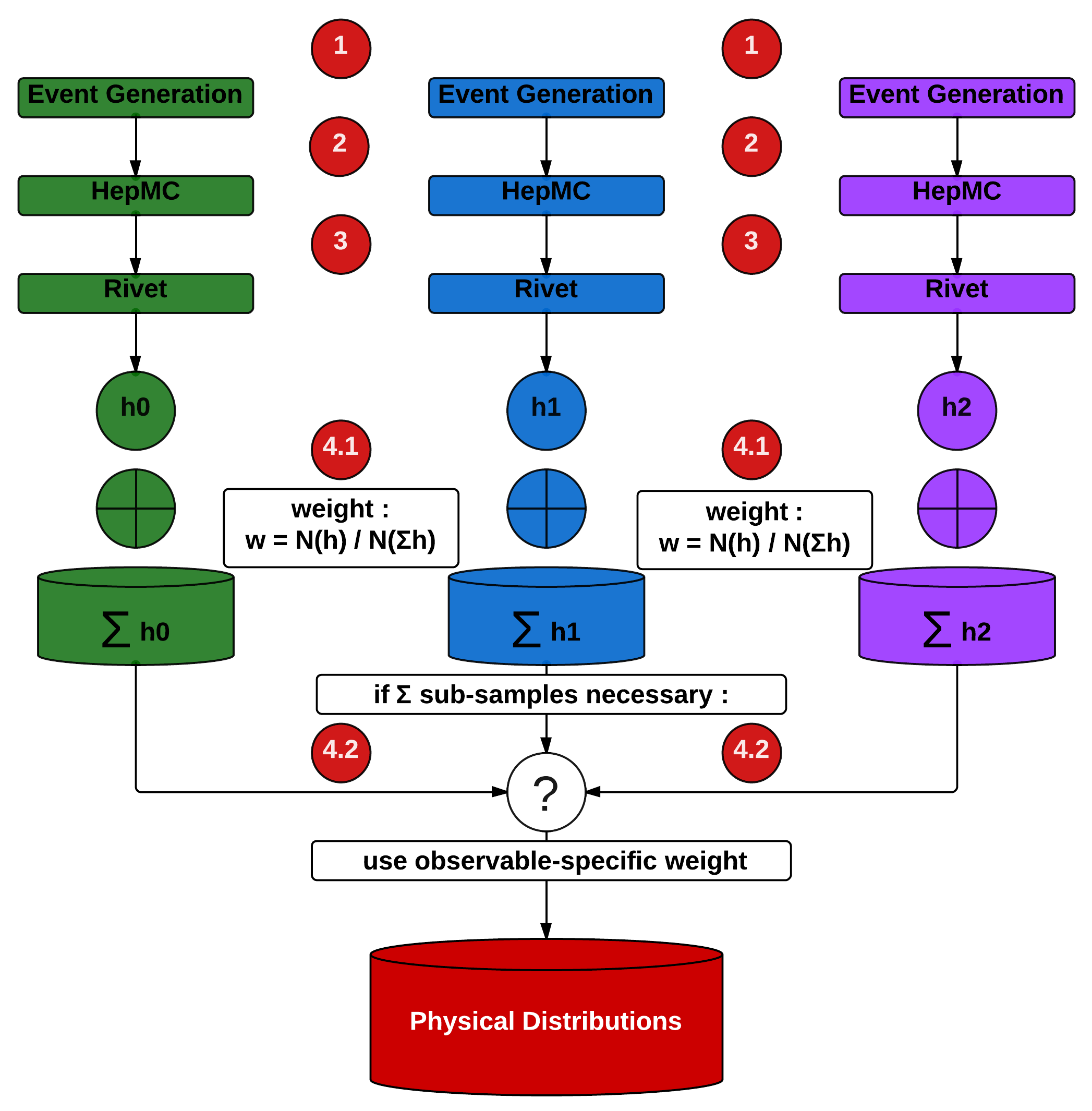}
\caption{A schematic representation of the path from the Event Generation to the Physical Distributions produced within the \mcplots framework. The sequential steps are labelled 1 -- 4. Their technical implementation in the \mcplots framework is described in \secRef{sec:newGenerator}. \label{Fig:generation_schema}}
\end{figure}
We explain the updates needed for implementation of the new generator supported by GENSER with the help of the
figure \figRef{Fig:generation_schema}. The figure shows the schematic 
representation of the path from the Event Generation to
the Physical Distributions produced within the \mcplots framework.
The sequential steps are labelled 1 -- 4. 

\subsection {Steps 1-3: from the Event Generation to \rivet histograms \label{subsec:steps1-3}}

Before running the event generation in step 1 of figure
\ref{Fig:generation_schema}, the generator 
configuration scripts and any other necessary files should be set up. 

The top-level wrapper script {\tt runAll.sh} (discussed in
\secRef{sec:updateGenerator}) contains the commands for multiple runs
of steps 1-3 of the \mcplots framework. Generator-specific code and
calls to generator-specific files for steps 1-2 are implemented in the
script: 
\vskip\vbspc\noindent\begin{minipage}{\textwidth}{\small
\begin{verbatim}
  $SCRIPTS/mcprod/rungen.sh . 
\end{verbatim}
}\end{minipage}\vskip\vbspc
The third step, executing \rivet, is not generator specific and is
handled by the script  
\vskip\vbspc\noindent\begin{minipage}{\textwidth}{\small
\begin{verbatim}
  $SCRIPTS/mcprod/runRivet.sh . 
\end{verbatim}
}\end{minipage}\vskip\vbspc
The {\tt runRivet.sh} relies on {\tt rungen.sh} and enables performing
steps 1-3 in one go.  

The script {\tt rungen.sh} handles calls of the generator-specific
files and provides parsing of the generator-specific parameters passed
to the upstream {\tt runAll.sh} and {\tt runRivet.sh}.  

As discussed in \secRef{sec:eventGeneration}, the physics
process-configuration files for each event generator should reside in the 
directory 
\vskip\vbspc\noindent\begin{minipage}{\textwidth}{\small
\begin{verbatim}
  $SCRIPTS/mcprod/configuration .
\end{verbatim}
}\end{minipage}\vskip\vbspc
\noindent
(See also the discussion of ``tune groups''
in \secRef{sec:tuneGroups}.) The location of the \code{\$SCRIPTS} directory was defined in
\secRef{sec:dirs}. 
Additional generator-specific files may be needed. Examples are linking
against the generator libraries of the GENSER project,
parsing the  generator configuration scripts, or providing 
generator-level event filters, used by some of the analyses in order
to enable efficient filling of the histograms. 
(Some concrete examples are given below.)
Such generator-specific files should be put into a dedicated directory
\vskip\vbspc\noindent\begin{minipage}{\textwidth}{\small
\begin{verbatim}
  $SCRIPTS/mcprod/(generator) .
\end{verbatim}
}\end{minipage}\vskip\vbspc
\noindent
Using \alp\ as an example, the \alp-specific files reside in the directory
\vskip\vbspc\noindent\begin{minipage}{\textwidth}{\small
\begin{verbatim}
  $SCRIPTS/mcprod/alpgen .
\end{verbatim}
}\end{minipage}\vskip\vbspc
\noindent

The process-specific files are passed
in text format using the standard \mcplots naming convention (``{\tt
  generator-process.params}''), for example: 
\vskip\vbspc\noindent\begin{minipage}{\textwidth}{\small
\begin{verbatim}
  alpgenherwigjimmy-winclusive.params
  alpgenpythia6-jets.params .
\end{verbatim}
}\end{minipage}\vskip\vbspc
\noindent
The parameters in these {\tt .params} files are grouped into sections
responsible for for various steps of event generation, using the
standard \alp formats:
\begin{myitemize}
\item parameters in sections {\tt [genwgt]} and {\tt [genuwgt]} are used for the generation of the weighted and unweighted events accordingly
\item the parameters used for MLM matching (see \cite{Mangano:2006rw}) are passed in section {\tt [addPS]}
\item any shower-specific parameters (e.g., tune selection) 
to be passed to the parton-shower generator are contained in {\tt [steerPS]}.
\item optionally, additional sections can be added as well.
\end{myitemize}
Examples of parameter files for $W$ + jets are available on the
\mcplots site.  
  
The code responsible for parameter parsing resides in the directory:
\vskip\vbspc\noindent\begin{minipage}{\textwidth}{\small
\begin{verbatim}
  $SCRIPTS/mcprod/alpgen/utils_alp
\end{verbatim}
}\end{minipage}\vskip\vbspc
\noindent
which also contains a dictionary of unweighting and MLM
matching efficiencies that enables the generation of a number 
of final unweighted \alp + parton-shower events as specified by the
user. 
  
The driver scripts for weighted and unweighted event generation reside
in the directory:
\vskip\vbspc\noindent\begin{minipage}{\textwidth}{\small
\begin{verbatim}
  $SCRIPTS/mcprod/alpgen/mcrun_alp
\end{verbatim}
}\end{minipage}\vskip\vbspc
\noindent
This directory also contains generator-level event filters, used by
some of the analyses in order to enable efficient filling of the
histograms. 

The files in the two directories just mentioned, 
{\tt utils\_alp/} and {\tt mcrun\_alp/}, are
used by the {\tt rungen.sh} script. The script supports
optional additional arguments beyond the standard {\tt \$mode \$conf} 
ones (described in \secsRef{sec:updateGenerator} and \ref{sec:mcGeneration}), 
so that extra generator-specific parameters can be included. In the case
of \alp, two extra parameters are currently given: the parton
multiplicity of the matrix element and inclusive/exclusive matching of
the given run.  (Tune-specific \alp parameters are passed 
at the time of the tune setup, see above. 
The event-generation scripts support the usage of parameters not
directly related to the tune.)
 
The step 2 in \figRef{Fig:generation_schema} corresponds
to  ensuring that the output of the event generation is provided in the
standard event record format: \hepmc. For many of the FORTRAN
generators, the conversion to the \hepmc event record, as well as
utilities for parameter settings and the event generation executable,
are provided by A Generator Interface Library, \agile~\cite{agile}. The C++
generators are generally able to handle these tasks directly. 
The \alp event record can be written in \hepmc format either
by invoking \agile using the unweighted events as inputs or by using
dedicated interface code following the \alp-internal parton shower
interface code examples and adopting the standard HEPEVT to \hepmc 
conversion utilities. The implementation of the FORTRAN
generators in the \agile framework in order to obtain the events in
the \hepmc format is not a prerequisite for running in the \mcplots
framework. It should however be noted that using \agile simplifies 
running of \rivet in step 3 in \figRef{Fig:generation_schema}. It is
therefore preferred over dedicated interfaces. 

\agile allows, by setting an input flag, to pass parton level events in
LHEF format~\cite{Alwall:2006yp} to the supported General Purpose
Event Generators. We 
thus anticipate that, when showering events in LHEF format with FORTRAN
event generators such as \py and \hw, no dedicated code will be
necessary for the steps 2 and 3 in figure \figRef{Fig:generation_schema}.

\subsection{Steps 4.1, 4.2: from single run histograms to Physical Distributions \label{subsec:CombiningResults}}

After the analysis run in step 3 in \figRef{Fig:generation_schema}, the \mcplots framework 
merges the single run histograms with the rest of the compatible analysis runs preceding it (steps 4.1 and 4.2).
In some cases the single run histograms already correspond to Physical Distributions and the merging is done in order to improve the statistical precision. 
In other cases several single run histograms need to be combined in order to obtain the Physical Distributions. For the histogram merging it is assumed that the histograms scale linearly with the number of events. 

The merging code is located in:
\vskip\vbspc\noindent\begin{minipage}{\textwidth}{\footnotesize
\begin{verbatim}
  $SCRIPTS/mcprod/merge .
\end{verbatim}
}\end{minipage}\vskip\vbspc
\noindent 
Once a new source histogram (h0, h1 or h2) in \figRef{Fig:generation_schema} is obtained, it should be merged to a common destination with the prior compatible histograms in the database. The run of the merging script:
\vskip\vbspc\noindent\begin{minipage}{\textwidth}{\footnotesize
\begin{verbatim}
  $SCRIPTS/mcprod/merge/merge.sh [source] [destination]
\end{verbatim}
}\end{minipage}\vskip\vbspc
\noindent 
handles the merging of the ASCII histogram inputs as well as the book-keeping of the event multiplicities and the sample cross-section. The merging code runs in two steps as follows:
\begin{itemize} 
\item first histograms that populate exactly the same regions of phase-space are merged in order to increase the statistical precision of the results. This is denoted as step 4.1 in \figRef{Fig:generation_schema}.
\item In the second pass of the code, the use-cases where a number of distinct generation runs are best suited to populate the phase-space needed for the final histograms are handled. This is denoted as step 4.2 in \figRef{Fig:generation_schema}. This step is omitted in cases where a single generation run populates the whole phase-space. 
\end{itemize}
The source histograms are stored in a directory and name structure, from which the analysis and generator setup can be inferred. The exact paths are specified in the {\tt runRivet.sh} script and contain the entries of the {\tt rivet-histograms.map} described in \secRef{sec:eventGeneration} as well as the generator, version, tune and any generator-specific parameters passed to the \mcplots scripts. The source file path and name thus provide the analysis and generator information to the merging script. This information is used in the script to decide which of the steps to perform and other merging details (detailed below). 

An example use-case that requires only one pass of the merging code
is, the standard one of combining multiple runs with different
random-number seeds but otherwise identical
settings.

An example use-case of the run that requires the second pass of the
merging code is filling of the histograms for the production of
$W$/$Z$ bosons in association with jets at the LHC, in the context of
multi-leg matrix-element matched samples. 
The physics distributions of interest have been measured up to large high-$p_T$ jet multiplicities by the LHC experiments~\cite{Aad:2010ab,Chatrchyan:2012jra,Chatrchyan:2013tna,Aad:2013ueu,Aad:2013vka,Chatrchyan:2013rla,Aad:2013ysa,Chatrchyan:2013jya}. 
The multi-leg generators such as \alp~\cite{Mangano:2002ea},
\textsc{Helac}~\cite{Cafarella:2007pc},
\mg~\cite{Alwall:2007st,Alwall:2011uj},
\sherpa~\cite{Gleisberg:2008ta}, and \textsc{Whizard}~\cite{Kilian:2007gr} 
are well suited for the physics
case. For \alp and \mg, the efficient population of phase-space needed for prediction of high extra jet multiplicities is frequently obtained by producing separate runs with fixed number of extra partons from the matrix element ({\tt Np}) that are passed through the parton shower and hadronization. In this way, the high {\tt Np} sub-samples which are more probable to populate the high extra jet bins in the physics events can be produced with larger integrated luminosity than the low  {\tt Np} respectively. A concrete example is the measurement of $W$ + jets by the ATLAS collaborating \cite{Aad:2010ab}, where the cross-section decreases by an order of magnitude per extra jet while the differential measurement is available for up to $\geq$ 4 extra jets. Another example requiring the second pass of the merging code is also a combine the physics distributions from generator runs that populate different regions of phase space using generator-level cuts or filters. This is a possible scenario for any generator. The merging code and the histogram structure described in \secRef{sec:hisFormat} is suited to address such use-cases.

In \figRef{Fig:generation_schema} the event generation branches in which the histograms h0, h1, h2(\ldots) are produced correspond to the sub-samples populating independent phase-space. An example are sub-samples with 0, 1, 2(\ldots) extra partons in the case of \alp or \mg in which case the exclusive matching criterion must be used for all but the highest multiplicity sub-sample which is matched inclusively. Multiple sub-samples could also be produced with different phase-space cuts to enhance the number of events in the target corners of the phase-space, such as cuts on the transverse momentum of the hard process partons, such that the high-$p_T$ events can be produced with higher integrated luminosity than the low-$p_T$ events. The merging code contains a procedure to detect and correctly sum the resulting histograms as illustrated in \figRef{Fig:generation_schema}.

The correct merging procedure of the histogram data depends on the normalization already assumed in the \rivet routines. In practice all the current use-cases could be addressed by adding two merging modes  {\tt ALPGEN\_XSECT} and {\tt ALPGEN\_FIXED} to the merging machinery. The {\tt ALPGEN\_XSECT} mode is used for histograms from \rivet runs, when no 
normalization is performed by \rivet. The {\tt ALPGEN\_FIXED} mode is used for histograms that are normalized according to the cross-section by \rivet. The new histograms from the run with {\tt Np} partons is merged with the existing histograms for the {\tt Np} sub-sample. The weights for merging of all the {\tt Np} samples needed to form the generator prediction for 
the observable are than evaluated such that all the {\tt Np} sub-samples are normalized to the same integrated luminosity in {\tt ALPGEN\_XSECT}, or with unit weights for {\tt ALPGEN\_FIXED} mode correspondingly. The choice of the correct merging procedure is implemented in the merging script {\tt merge.sh} and relies on the histogram naming conventions. In particular the entries of {\tt rivet-histograms.map}, that are constituents of the histogram path, can serve to assign the correct merging mode according to the analysis and generator details. For example the choice of the merging procedure for currently implemented \alp runs uses the \code{Obs} and \code{process} fields of {\tt rivet-histograms.map}~(\secRef{sec:eventGeneration}).

The histogram METADATA (e.g. cross-section and the number of events) described in \secRef{sec:hisFormat} is consistently updated with the merged values. In addition the sub-sample specific fields are added in the format:
\vskip\vbspc\noindent\begin{minipage}{\textwidth}{\footnotesize
\begin{verbatim}
  samples_X=X_h0:X_h1:X_h2 .
\end{verbatim}
}\end{minipage}\vskip\vbspc
\noindent
Here {\tt X} denotes a quantity, e.g. cross-section. The {\tt X\_h} denote the value of the quantity for the individual sub-sample. Hence merging the {\tt Np0, Np1, Np2} sub-samples in {\tt ALPGEN\_FIXED} mode would result in the following METADATA for {\tt X=crosssection}~(in [pb]):
\vskip\vbspc\noindent\begin{minipage}{\textwidth}{\footnotesize
\begin{verbatim}
  crosssection=25554.1
  samples_crosssection=20831.403148:4285.7315005:436.96532523
\end{verbatim}
}\end{minipage}\vskip\vbspc
\noindent

It should be noted that the merging in the \mcplots needs to proceed
on-the-fly, since new samples are produced continuously. This makes
the merging more challenging than in standard case, where the final
statistics and cross-sections of all the {\tt Np} samples is known at
the time of analysis and plot production. Thus, in the standard case, 
the {\tt Np} samples can be normalized to the same integrated
luminosity prior to the plot production, while the \mcplots machinery
needs to deal with merging of the already produced plots. 

In case the need arises, further merging modes could be added to the merging structure. For this it is however crucial, that the information needed for consistent merging is available after the \rivet run and correctly transferred to the merging routine. An example piece of such information is the cross-section in the MLM-matching applications, where the final cross-section is only known after the parton shower. Hence, in case the cross-section is not correctly transferred from the matrix-element to the parton-shower generator, the automatic extraction of the information needed for the consistent merging would fail.

\section{LHC@home 2.0 and the Test4Theory Project\label{sec:t4t}}
\lhcathomeclassic\ started as an outreach project for CERN’s 50th Anniversary in
2004. The project calculated the stability of proton orbits in the LHC
ring, using a software called
\sixtrack~\cite{RobertDemolaize:2005iq}, distributed to volunteers via
the Berkeley Open Infrastructure for Network Computing
(BOINC)~\cite{boinc}, a popular middleware for volunteer computing.  
When \mcplots\ was initially conceived, it was clear that significant
sustained computing resources would be needed, and it was natural to
consider if a setup similar to that of \sixtrack\ could be created to
meet those needs. The three main reasons were: 
1) the computing resources envisaged for \mcplots\ would have been 
comparable to (or larger than) the total amount of computing power
then available to the CERN theory group;  
2) the developers of \lhcathomeclassic\ were keen to explore possibilities to
expand the volunteer computing framework towards HEP physics
simulations; 3) we saw event simulations as providing a natural way to involve
the public in doing LHC science, without the complications that would
have accompanied analysis of real data. 

A major challenge for the \sixtrack\ project had been the heterogenous
nature of the resources provided by volunteers. In particular, it is 
mandatory to support Windows platforms, which the HEP 
scientific-computing community is significantly less familiar with 
than variants of Linux, UNIX, or Mac OSX. In the context of event-generator 
simulations, we viewed the time-consuming task of porting and maintaining our
code over such a large range of platforms as a showstopper. An elegant
solution to this problem, which factorizes the IT issues almost
completely from the scientific software development, is
virtualization. The development at CERN of 
a Scientific-Linux based Virtual-Machine architecture (\cernvm) along
with a generic and scalable infrastructure for integrating it into 
cloud-computing environments (\copilot) were the two main innovations
that allowed us to start the \tft\ project, which now provides the 
computing backbone for \mcplots. This represented the first
virtualization-based volunteer computing project in the world, and, 
with the new additions, the name was updated to \lhcathome. 

\cernvm itself can be installed using any of a number of different
so-called virtualization hypervisors (virtual-machine host software),
most of which have been designed to add very little overhead to the
virtualized simulation. One that is open-source and available on all 
platforms is \textsc{virtualbox}, which thus is a prerequisite for
running simulations for \tft. 

The file system of \cernvm\ is designed so that only files that are
actually accessed are downloaded to the local disk; a huge virtual
library can therefore in principle be made available, without causing
a large local footprint or long download times. 

At the time of writing, the \lhcathome\ project still uses BOINC as
the main middleware for distributing computing jobs to
volunteers. What is different with respect to the older
\sixtrack\ project is that the  BOINC tasks in \tft are merely
wrappers for a VM that is started up on the remote computer. Each such
VM task has a default lifetime of 
24 hours, since volunteers only gain ``BOINC credits'' each time a
BOINC task completes. Inside the wrapper, the VM itself constantly 
receives jobs, processes them, and sends
output back to \mcplots. This communication is handled 
by \copilot. 

Alpha testing of the system internally at CERN began in October
2010. This first test phase was quite technical, concentrating on the
requirements on the virtual-machine architecture, on the stability and
steady supply of jobs, and on development of the simulation packages
themselves. During 2011, a small number of external volunteers were
gradually connected as well, many of whom participated actively in
testing and debugging this first edition of the system. 
By the end of the alpha test stage, in July 2011, the system
operated smoothly and continously with about 100 machines connected
from around the globe. 

During beta testing, in the latter half of 2011, 
the main line of attack was the scalability of
the system. In a first beta trial in August 2011, the system was
opened up to the broader public in combination with a press release
from CERN. This resulted in such a massive amount of new subscriptions
that the system then in place could not handle it, resulting in
crashes. A procedure was then introduced by which 
volunteers wishing to participate could sign up for participation
codes which were issued incrementally. This gradually brought the number of
connected participants up to around 2500, while we were able to
monitor and improve the scalability significantly. During its second phase, the
participation-code restriction was removed, and the number of
successfully connected hosts increased to about 6500
machines by the end of the beta trial (with a noticeable spike around
the CERN Press Release on Dec 13, 2011 concerning the possible hints
of a Higgs boson; another significant spike was seen on July 4th 2012,
when the discovery was confirmed). 

\begin{figure}[t]
\centering
\includegraphics[scale=0.55]{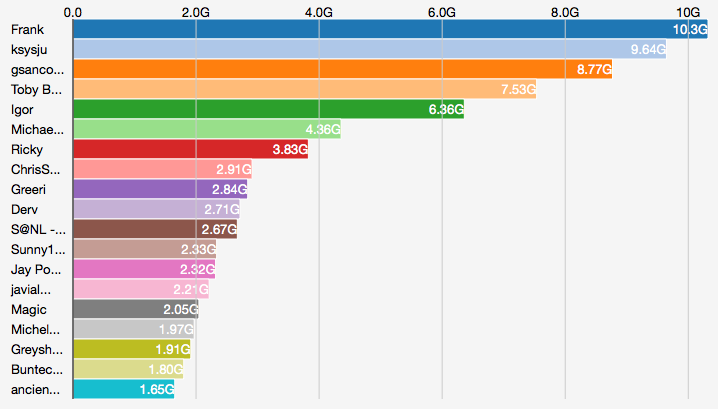}
\caption{\tft Leaderboard (top contributors in number of events
  generated), as of May 2013. The up-to-date
  online version can be consulted
  at~\cite{test4theory}.\label{fig:leaderboard}}  
\end{figure}
The system is now fully operational, with several thousand volunteers
donating CPU cycles on a daily basis, and individual volunteers
already having generated billions of events each, 
cf.\ the {\sl Test4Theory Leaderboard} reproduced in \figRef{fig:leaderboard}. 
At the time of writing, in June 2013, the number of
volunteers who have completed at least one work package is 8000,
with a combined total of roughly 13500 machines. The number of
instantaneously connected machines fluctuates between 600 - 800.

These contributions have made a
crucial difference in being able to generate the massive amounts of
statistics required for the innumerable combinations of generators,
tunes, and experimental analyses provided on \mcplots.

\section{Conclusions}

In this paper, we have provided an elementary user guide for the
\mcplots\ site,
and summarized its technical implementation. We intend the site to be 
broadly useful to the particle physics community, as a resource for MC
validation and tuning, and as an explicit browsable 
repository of \rivet analyses. It also provides a possibility for the
public to engage in the scientific process, via the \lhcathome project
\tft. 

In the near future, we plan to
extend the site by adding more possibilities for users to create their
own plots and control how they look. We will also aim
to provide event-generator authors with a pre-release validation
service, in which a would-be new version of a generator or tune
parameter card can be uploaded securely and a private version of the
site made available on which the corresponding results can be compared
with the standard set of generators and tunes. 

Plans for future
developments of \tft include 
minimizing the size of the downloadable \cernvm
image for \tft ($\mu$\cernvm), a new delivery method for \tft that
would allow one to download, install, and run \tft directly from a web
browser (\tft {\sl Direct}), and the development of a new interactive
citizen-science application based on the \tft framework.  

Note to users of \mcplots: we are of course very grateful if a
reference to this work is included when using content from \mcplots, 
but even more importantly, we ask our users to please acknowledge 
the original sources for the data, analyses, physics models, and 
computer codes that are displayed on \mcplots. We have tried to make
this as easy as 
possible, by including links to the original experimental papers
together with each plot. Other references that may be appropriate,
depending on the context, include \hepdata~\cite{Buckley:2010jn}, 
\rivet~\cite{Buckley:2010ar}, MC generator
manuals~\cite{Bopp:1998rc,Mangano:2002ea,Sjostrand:2006za,Sjostrand:2007gs,Bahr:2008pv,Gleisberg:2008ta,Werner:2010aa},
and relevant physics and/or tuning papers. 

As an example of good practice, and to ensure maximal clarity and
reproducibility, all plots on \mcplots include explicit generator names and
version numbers for all curves appearing on the plot, together with
the relevant experimental reference. When combining an 
ME generator {\sl X} with a (different) shower generator {\sl Y}, 
both names and version numbers are shown explicitly.
This is not only to
give proper credit to the authors, but since the physics
interpretation of the calculation depends on how it was
performed. 

\subsection*{Acknowledgements}
We are indebted to many members of 
the Monte Carlo community for their feedback and help with validations
of the generators and analyses included on \mcplots.
Many short-term students, visitors, and fellows have
contributed with 
\rivet analyses and other content to the site, in particular
D.~Konstantinov (the \mcplots plotting tool), A.~Pytel (server
infrastructure and the MC validation view), S.~Sen (forward LHC
analyses in \rivet and on \mcplots, and implementation of the \epos
and \pho generators), and J.~Winter (the $t\bar{t}$ MC
analyses).   
We thank J.~Katzy, for carefully reading this manuscript and giving 
feedback on the \mcplots site, M.~Mangano, for his strong support of
this project from its inception to completion and for help with the
\alp implementation, and T.~Pierog, for help with the \epos implementation.

We express our profound admiration of the teams responsible
for the development of \rivet, \cernvm, and \lhcathome, without
which the \tft project would have been impossible. Thanks
also to the many volunteers who have provided computing  
resources, especially to those who have taken active part
in the alpha and beta testing phases of \tft. 
We also 
thank the CERN PH-SFT group, in particular the Generator Services
(GENSER) team, and the CERN IT
Division. 

This work was supported in part by the LHC Physics Center at CERN
(LPCC), by the Marie Curie research training
network ``MCnet'' (contract number MRTN-CT-2006-035606),
and by the National Science Foundation
under Grant No.\ NSF PHY11-25915.

\bibliographystyle{utphys}
\bibliography{mcplots-arxiv}

\end{document}